\documentclass{aa}

\usepackage{graphicx}
\usepackage{txfonts}
\usepackage{amssymb}
\usepackage{amsmath}
\usepackage{graphicx}
\usepackage[colorlinks=true, allcolors=blue]{hyperref}
\usepackage{booktabs}
\usepackage{amsfonts}     
\usepackage[T1]{fontenc} 
\usepackage[utf8]{inputenc} 
\usepackage{multicol}
\usepackage{xcolor}
\usepackage[switch]{lineno}
\usepackage{gensymb}
\usepackage{soul}
\usepackage{rotating}
\usepackage{comment}
\usepackage{multirow}
\usepackage{longtable}
\usepackage{lscape}
\usepackage{makecell}

\newcommand{\eqb}{\begin{eqnarray}}
\newcommand{\eqe}{\end{eqnarray}}

\newcommand{\subsubsubsection}[1]{\paragraph{#1}\mbox{}\\}
\begin{document}

\title{Optimizing the hunt for extraterrestrial high-energy\\ neutrino counterparts}

\subtitle{}

\author{
Pouya M. Kouch \inst{\ref{UTU},\ref{FINCA},\ref{MRO}} \thanks{\href{mailto:pouya.kouch@utu.fi}{pouya.kouch@utu.fi}} \orcid{0000-0002-9328-2750},
Elina Lindfors \inst{\ref{UTU},\ref{FINCA}} \orcid{0000-0002-9155-6199}, 
Talvikki Hovatta \inst{\ref{FINCA},\ref{MRO}} \orcid{0000-0002-2024-8199}, 
Ioannis Liodakis \inst{\ref{IA_FORTH_Crete},\ref{FINCA},\ref{NASA_Alabama}} \orcid{0000-0001-9200-4006}, 
Karri I.I. Koljonen \inst{\ref{NTNU}} \orcid{0000-0002-9677-1533}, 
Kari Nilsson \inst{\ref{FINCA}} \orcid{0000-0002-1445-8683}, 
Jenni Jormanainen \inst{\ref{UTU},\ref{FINCA}} \orcid{0000-0003-4519-7751}, 
Vandad Fallah Ramazani \inst{\ref{FINCA}} \orcid{0000-0001-8991-7744}, 
Matthew J. Graham \inst{\ref{CalTech}} \orcid{0000-0002-3168-0139} 
}

\institute{
Department of Physics and Astronomy, University of Turku, FI-20014 Turku, Finland \label{UTU}
\and
Finnish Centre for Astronomy with ESO (FINCA), Quantum, Vesilinnantie 5, University of Turku, FI-20014 Turku, Finland \label{FINCA}
\and
Aalto University Mets\"ahovi Radio Observatory, Mets\"ahovintie 114, FI-02540 Kylm\"al\"a, Finland \label{MRO}
\and
Institute of Astrophysics, FORTH, N. Plastira 100, GR-70013 Heraklion, Greece \label{IA_FORTH_Crete}
\and
NASA Marshall Space Flight Center, Huntsville, AL 35812, USA \label{NASA_Alabama}
\and
Institutt for Fysikk, Norwegian University of Science and Technology, H{\o}gskloreringen 5, Trondheim, 7491, Norway \label{NTNU}
\and
Division of Physics, Mathematics and Astronomy, California Institute of Technology, Pasadena, CA91125, USA \label{CalTech}
}

\date{Received December 03, 2024; accepted February 23, 2025}

 
\abstract{It has been a decade since the IceCube collaboration began detecting high-energy (HE) neutrinos originating from cosmic sources. Despite a few well-known individual associations and numerous phenomenological, observational, and statistical multiwavelength studies, the origin of astrophysical HE neutrinos largely remains a mystery. To date, the most convincing associations link HE neutrinos with active galactic nuclei (AGNs). Consequently, many studies have attempted population-based correlation tests between HE neutrinos and specific AGN subpopulations (such as blazars). While some of the associations are suggestive, no definitive population-based correlation has been established. This could result from either a lack of a population-based correlation or insufficient detection power, given the substantial atmospheric neutrino background. By leveraging blazar variability, we performed spatio-temporal blazar-neutrino correlation tests aimed at enhancing detection power by reliably incorporating temporal information into the statistical analysis. We used simulations to evaluate the detection power of our method under various test strategies. We find that: (1) with sufficiently large source samples, if 20\% of astrophysical HE neutrinos originate from blazars, we should robustly observe $\sim$4$\sigma$ associations; (2) a counting-based test statistic combined with a top-hat weighting scheme (rather than a Gaussian one) provides the greatest detection power; (3) applying neutrino sample cuts reduces detection power when a weighting scheme is used; and (4) in top-hat-like weighting schemes, low p-values do not occur arbitrarily with an increase in the HE neutrino error region size (any such occurrence is indicative of an underlying blazar--neutrino correlation).}


\titlerunning{Optimizing the hunt for extraterrestrial high-energy neutrino counterparts}
\authorrunning{Kouch et al.}

\maketitle
%

\section{Introduction} \label{sec_intro}
In 2008, the IceCube Neutrino Observatory began observing the sky for high-energy (TeV--PeV) neutrinos (hereafter "neutrinos"; e.g., \citealt{aartsen2013_syserr_1deg, IC2021_history_ref}). Since then, hundreds of such neutrinos have been detected from cosmic sources (\citealt{IC2013_detection_of_astrophysical_neut}). Despite more than a decade of observations, the origin of these astrophysical neutrinos remains elusive. This uncertainty is partly due to their intrinsic rarity, and partly due to observational challenges posed by the atmospheric neutrino background (e.g., \citealt{IC2021_review,troitsky2021}).

Active galactic nuclei (AGNs), powered by supermassive black holes, have long been suspected to be sources of neutrinos (e.g., \citealt{mannheim89, Stecker1991_AGN_neut, mannheim1992, mucke01, becker2008, Dermer2014_blz_neut_connection, Stecker2013_AGN_neut, hooper_kathryn2023_leptonic_neut_production}). A fraction of AGNs exhibit highly collimated jets of relativistically moving plasma. When the jet of an AGN aligns with our line of sight, the AGN is classified as a blazar. The small viewing angle of the jet causes all emission from the relativistically accelerated plasma to become extremely Doppler-boosted, making blazars some of the brightest objects in the Universe (for recent reviews, see, e.g., \citealt{Bottcher19,blandford2019,hovatta_lindfors2019}).

In 2017, the IceCube Neutrino Observatory detected a high-energy neutrino coming from the direction of the blazar TXS~0506+056, which was in a state of heightened electromagnetic activity ($>$3$\sigma$ spatio-temporal chance coincidence; \citealt{IC2018_txs0506}). Subsequent analysis of archival data revealed an excess of neutrinos from the direction of TXS~0506+056. Later, the IceCube collaboration identified another AGN, the Seyfert-II NGC~1068, as a source of neutrinos with a $4.2\sigma$ significance level (\citealt{icecube_collab2022_ngc1068}). Additionally, two other blazars, PKS~1424+240 and GB6~J1542+6129, were identified as potential neutrino emitters (\citealt{IC2021_pks1424_rg6j1542}).

While many blazar--neutrino correlation studies have focused on individual source associations (e.g., \citealt{krauss2014, kadler2016, righi2019, Franckowiak2020, Das2022_TXS0506}), many studies have also investigated the blazar--neutrino connection via population-based statistical analyses (e.g., \citealt{aartsen2020, plavin2020, giommi2020, plavin2021, smith2021, hovatta2021, zhou2021, bartos2021, kun2022, buson2022, plavin2023, novikova2023, buson2023, bellenghi2023_5bzcat_and_rfc_expansion, suray2024, Lu2024, plavin2024_xray_connection, Kouch2024_CGRaBS_update}), aiming to leverage the abundance and precise sky localization of blazars to enhance detection power. Several of these studies found hints of a potential connection, but none conclusively established a correlation between blazar populations and neutrinos. Therefore, it is crucial to identify the most effective detection strategies, which is the focus of this work.

An important feature of blazars is the stochastic nature of their energy output. Blazars exhibit significant variability across nearly all electromagnetic bands, alternating between quiescent and enhanced emission states (commonly referred to as "flaring" periods). In our previous studies (\citealt{hovatta2021} and \citealt{Kouch2024_CGRaBS_update}; hereafter H21 and K24, respectively), we focused on flaring periods in the lower energy bands. In the radio band, flares often trace the enhancement of the overall jet activity (e.g., \citealt{Lahteenmaki2003_radio_gamma_correlation, leon_tavares2011_radio_gamma_correlation}), suggesting that major flares could correlate with enhanced neutrino production (e.g., \citealt{plavin2020}, H21). In the optical band, the variability typically traces the slower radio-band variability as well as the faster X-ray and $\gamma$-ray flares (e.g., \citealt{Lindfors2016_radio_optical_correlation}), which requires efficient particle acceleration and cooling (e.g., \citealt{oikonomou2019_indepth_blz_neut_connection, kreter2020_fermi_flare_v_neutrinos, stathopoulos2022_xray_flare_v_neutrinos}). This is the basis for the hypothesis of a temporal connection between flaring periods and neutrino emission. 

Incorporating the temporal information of neutrinos into blazar--neutrino correlation schemes is crucial for boosting detection power (e.g., \citealt{Abbasi2024_time_dependent_analysis_vs_MOJAVE}), especially given the rather poor sky localization of neutrinos. This temporal connection between blazar flares and neutrino production may also imply a nontemporal effect: if more neutrinos are produced during flares, then a more frequently flaring blazar is more likely to be a neutrino emitter on average. Additionally, since $\gamma$-ray-emitting blazars are more variable in both radio and optical bands than $\gamma$-ray-dark blazars (e.g., \citealt{Richards2011_radio_vs_gamma, Richards2014_var, Hovatta2014_opt_vs_gamma}), we could likewise expect neutrino-emitting blazars to show greater time-averaged variability than neutrino-dark blazars. These are admittedly rather simplistic interpretations of the complex flaring behavior in blazars. Nevertheless, we consider it worthwhile to investigate a possible connection between time-averaged variability and neutrinos (see Sect. \ref{sec_data_blz}).

In our previous simulations (\citealt{liodakis2022_wild_hunt}, hereafter L22), we evaluated the feasibility of detecting a significant spatio-temporal blazar--neutrino correlation using available data ($\sim$1800 blazar radio light curves and 56 IceCube high-energy neutrino events, as in H21). We showed that incorporating the temporal dimension is necessary for establishing such a correlation reliably. We also found that increasing the blazar sample size strengthened the correlation. In this study, in preparation for our upcoming blazar--neutrino correlation study set to utilize $\sim$4000 blazar optical light curves, we performed similar simulations with an expanded sample of blazar light curves ($\sim$4000; see Sect. \ref{sec_data_blz}) and a larger, better-sampled catalog of IceCube high-energy neutrinos (283 events as used in K24, mostly taken from IceCat-1; \citealt{abbasi2023_cat1}; see our Sect. \ref{sec_data_neut}). Our focus is to simulate blazars with an underlying spatio-temporal correlation with neutrinos, and then evaluate the effectiveness of different correlation test strategies to establish this induced connection. This allows us to determine the most optimal test strategies for application to observed datasets in our upcoming study.

We ran these tests on five blazar samples: two for sanity checks, two for special cases, and one for a potentially realistic scenario (see Sect. \ref{sec_tests_sim_sample}). In short, we first calculated the correlation strengths between neutrinos and a population of 4000 random blazars (the first sanity-check sample) across all test strategies. Then, we added a number (tens to hundreds) of spatio-temporally neutrino-associated blazars to this random sample and reevaluated the correlation strengths. We repeated this simulation process 1000 times to determine the fraction of significant detections produced by each test strategy; this allowed us to systematically identify the most effective correlation test strategies for detecting a potential spatio-temporal blazar--neutrino correlation.

In Sect. \ref{sec_data_neut} we describe the neutrino data used. In Sect. \ref{sec_data_blz} we present the blazar sample and variability measures. In the beginning of Sect. \ref{sec_tests}, we outline the basics of the spatio-temporal search method applied in H21, K24, and this study. In Sect. \ref{sec_tests_sim_sample} we explain how we simulated five blazar datasets with varying degrees of spatio-temporal connection to neutrinos. In Sects. \ref{sec_tests_TS_parameters}, \ref{sec_tests_handling_neutrinos}, and \ref{sec_tests_combining} we describe how we set up and combined different correlation test strategies. We present and discuss our results in Sect. \ref{sec_results}. Finally, in Sect. \ref{sec_conclusions} we summarize our findings.

\section{Data} \label{sec_data}

\subsection{High-energy neutrino events} \label{sec_data_neut}
The majority of the neutrino data used in this study are from the IceCube collaboration's IceCat-1 (\citealt{abbasi2023_cat1}), which recorded the first catalog of muon-track high-energy neutrino events between May 2011 and December 2022 based on real-time alert selection criteria. Our sample also includes events from an earlier IceCube collaboration study (\citealt{abbasi2022_extra_neut}), which analyzed events between May 2009 and December 2018; it adds 16 events due to an earlier start date (when the telescope was in partial configuration) and the inclusion of non-real-time events. Together, our combined dataset termed IceCat1+, contains a total of 283 events. We previously used IceCat1+ in K24, where it is available as an electronic table.

\begin{figure}
    \centering
    \includegraphics[width=0.49\textwidth]{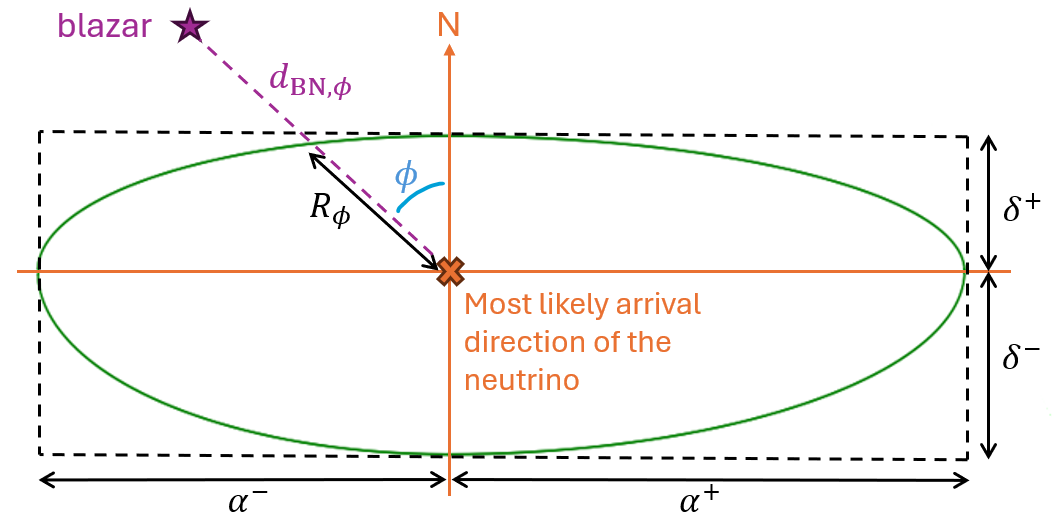}
    \caption{Spatial orientation of an example blazar--neutrino pair. The blazar is shown as a purple star. The most likely arrival direction of the neutrino is shown by the orange cross. The $\gtrsim$90\%-likelihood error region of the neutrino (see Sect. \ref{sec_data_neut}) is shown by the green ellipsoid. $d_{\mathrm{BN},\phi}$ is the distance between the blazar and the center of the neutrino error ellipsoid, shown by the dotted purple line. $R_\phi$ is the distance between the edge and center of the ellipsoid along the line that goes through the blazar, shown by the black double-headed arrow oriented at a $\phi$ angle from the northern axis. $\alpha^{+/-}$ and $\delta^{+/-}$ represent the asymmetric $2\sigma$ Gaussian error bars in RA and Dec., respectively.}
    \label{fig_blazar_neut_sketch}
\end{figure}

Each neutrino in IceCat1+ has an estimated kinetic energy and a "signalness" ($\mathcal{S}$; a measure indicating the likelihood of the neutrino being of astrophysical origin). Additionally, the IceCube collaboration provides maximum-likelihood contours on the sky, resembling ellipsoids, centered on the most probable arrival direction of each neutrino (e.g., Fig. 3 in \citealt{abbasi2023_cat1}). They report the right ascension (RA) and declination (Dec.) of these arrival directions along with four asymmetric error bars in the RA and Dec. directions, representing the 90\%-confidence-level contour size. The error bars are determined by bounding the 90\%-confidence-level contour within a rectangle. Since maximum-likelihood contours generally resemble ellipsoids with a Gaussian drop-off, it is mathematically convenient to approximate them using the four asymmetric RA and Dec. error bars entirely enclosing the 90\%-confidence-level contour area. We refer to this region as the $\gtrsim$90\%-likelihood error region (see Fig. \ref{fig_blazar_neut_sketch}). We note that each of the four asymmetric error bars is taken as a one-sided, $2\sigma$ Gaussian error.

In IceCat1+, the median signalness is $\widetilde{\mathcal{S}}=0.429$ and the median area of the $\gtrsim$90\%-likelihood error region is $\widetilde{\Omega}=6.63~\mathrm{deg}^2$. For each neutrino, the latter quantity was calculated as\begin{equation} \label{eqn_omega}
    \Omega = \frac{\pi}{4}\left(\alpha^+\cdot\delta^++\alpha^-\cdot\delta^++\alpha^-\cdot\delta^-+\alpha^+\cdot\delta^-\right)
,\end{equation}
where $\alpha^{+/-}$ and $\delta^{+/-}$ represent the asymmetric $2\sigma$ Gaussian error bars in the RA and Dec. directions, respectively (see Fig. \ref{fig_blazar_neut_sketch}).

\subsection{Blazars} \label{sec_data_blz}
As mentioned in Sect. \ref{sec_intro}, our goal is to perform simulations to test the detection power of various strategies in advance of our upcoming blazar--neutrino correlation study, which will utilize $\sim$4000 optical light curves. In line with this, we simulated blazar datasets containing around 4000 sources. The simulation details are given in Sect. \ref{sec_tests}. In this subsection we introduce the blazar population used and their variability measures. 

Decades of studies on the variability of blazars in the optical band have revealed that these sources exhibit significant differences in timescale, amplitude, and duty cycle (see Sect. \ref{sec_intro}). To parameterize both time-averaged and time-resolved variability, we used CAZ (CRTS+ATLAS+ZTF) optical light curves of blazars from the CGRaBS sample, as defined in K24. These light curves combine data from several all-sky surveys and, where available, from dedicated monitoring programs such as the Katzman Automatic Imaging Telescope (KAIT\footnote{\href{https://w.astro.berkeley.edu/bait/kait.html}{https://w.astro.berkeley.edu/bait/kait.html}}; \citealt{filippenko2001, liodakis2018_cross_correlation1}) and Tuorla\footnote{\href{https://users.utu.fi/kani/1m/}{https://users.utu.fi/kani/1m/}} (\citealt{nilsson2018}). The all-sky surveys include the Catalina Real-time Transient Survey (CRTS\footnote{\href{https://crts.caltech.edu/}{https://crts.caltech.edu/}}; \citealt{drake2009}), the Asteroid Terrestrial-impact Last Alert System survey (ATLAS\footnote{\href{https://fallingstar-data.com/}{https://fallingstar-data.com/}}; \citealt{tonry2018}), and the \textit{Zwicky} Transient Facility survey (ZTF\footnote{\href{https://irsa.ipac.caltech.edu/Missions/ztf.html}{https://irsa.ipac.caltech.edu/Missions/ztf.html}}; \citealt{bellm2019}). We compiled the light curves by performing forced-photometry on the sky coordinates of the CGRaBS blazars. While the CGRaBS sample is complete down to a 4.8~GHz flux density of 65~mJy and radio spectral index $\alpha > -0.5$ where S $\propto \nu^{\alpha}$, it over-represents low synchrotron peaked blazars, which are brighter than other blazar subclasses in the radio band. Thus, while the CGRaBS sample may not fully represent the entire blazar population, it provides a suitable basis for the general parameterization of their variability.

For the time-averaged variability measure, we used fractional variability ($F_\mathrm{var}$), which quantifies the overall variability of the flux density over a specific period. For a sequence of data points, $x_i \pm \epsilon_i$, where $i=1, \dots, N$, fractional variability is defined as follows:
\begin{equation} \label{eqn_Fvar}
    F_\mathrm{var} \equiv \sqrt{\frac{S^2-\overline{\epsilon}^{~2}}{\overline{x}^{~2}}}
,\end{equation}
where $\overline{x}=\frac{1}{N} \sum_{i=1}^{N}x_i$ is the sample mean, $S^2= \frac{1}{N-1} \sum_{i=1}^{N}(\overline{x}-x_i)^2$ is the sample variance, and $\overline{\epsilon}^{~2}=\frac{1}{N}\sum_{i=1}^{N} \epsilon_i^{2}$ is the mean square error of the sample (e.g., \citealt{Gliozzi2003_Fvar, Vaughan2003_Fvar, MAGIC2024_mrk501_Fvar}).

For the time-resolved variability measure, we used the activity index (AI), which we define here as the flux density at a given time ($S_t$) normalized by the median flux density ($\widetilde{S}$). We note that this definition differs from that in H21 and K24, where $S_t$ was taken as the average flux density within a time window, while here $S_t$ is calculated instantaneously at time $t$.

\begin{figure}
    \centering
    \includegraphics[width=0.49\textwidth]{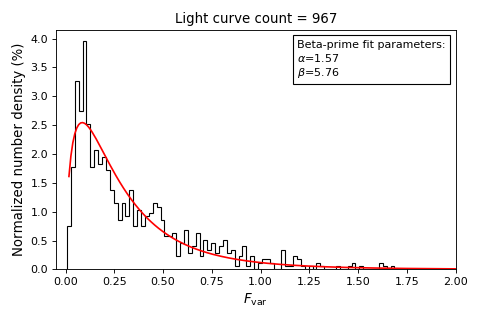}
    \caption{Global distribution of $F_\mathrm{var}$ for 967 CAZ light curves of the CGRaBS sample. The red curve shows a Beta-prime parameterization of the distribution. The Beta-prime fit parameters are $\alpha=1.57$ and $\beta=5.76$.}
    \label{fig_global_Fvar}
\end{figure}

To ensure statistical robustness, we limited our sample to light curves with at least 100 data points, resulting in 967 CAZ light curves. Figure \ref{fig_global_Fvar} shows the global distribution of $F_\mathrm{var}$ for these light curves parametrized using a Beta-prime distribution with a probability density function $f(x)=x^{\alpha -1} (1+x)^{-\alpha - \beta}/\mathcal{B}(\alpha,\beta)$ where $\mathcal{B}(\alpha, \beta)$ is the Beta function. The Beta-prime function is suitable for this parameterization as it is continuous and covers the positive real line (e.g., \citealt{Blinov2016_beta_func,Hovatta2016_beta_func,Bourguignon2018_betaprime}). The best-fit Beta-prime parameters for the $F_\mathrm{var}$ distribution are $\alpha=1.57$ and $\beta=5.76$.

Regarding AI, a flux density value around time $t$ is typically chosen (e.g., H21 and K24). However, for computational efficiency, we quantified AI directly from the flux density ($S$) distribution of each blazar (see Sect. \ref{sec_tests_sim_sample_random}). Thus, for each simulated blazar, we generated a random, yet realistic, flux density distribution.

\begin{figure}
    \centering
    \includegraphics[width=0.49\textwidth]{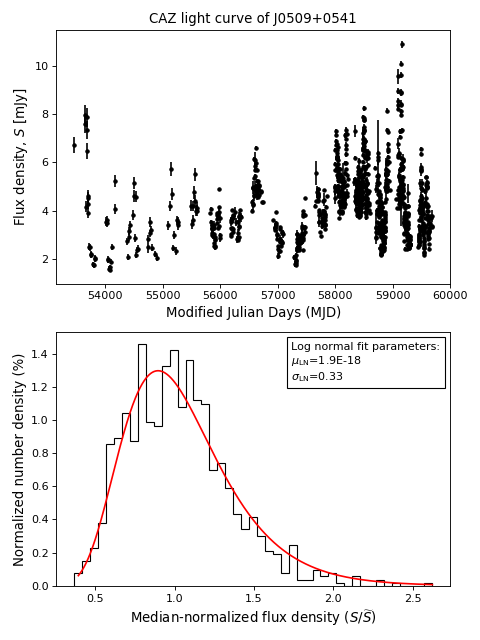}
    \caption{CAZ optical light curve of the blazar J0509+0541 (also known as TXS~0506+056). \textit{Top}: CAZ light curve. \textit{Bottom}: Corresponding median-normalized flux density ($S/\widetilde{S}$) distribution and its best-fit lognormal distribution (shown as the red curve with the parameters $\mu_\mathrm{LN} \approx 10^{-18}$ and $\sigma_\mathrm{LN}=0.33$).}
    \label{fig_LC_CAZJ0509+0541}
\end{figure}

\begin{figure}
    \centering
    \includegraphics[width=0.49\textwidth]{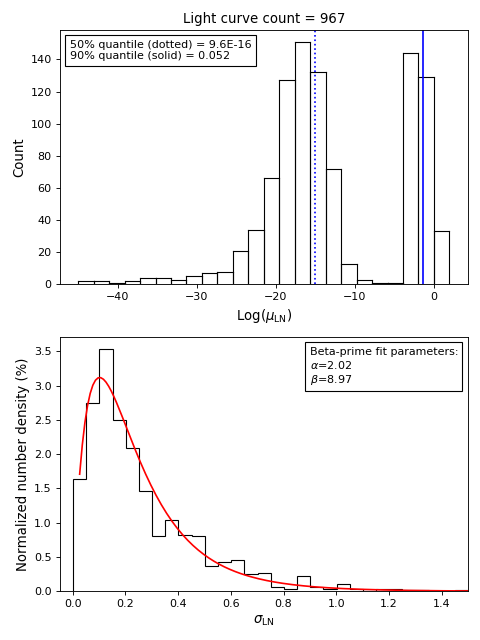}
    \caption{Global distribution of the lognormal best-fit parameters ($\mu_\mathrm{LN}$ and $\sigma_\mathrm{LN}$) on all 967 CAZ flux density distributions. \textit{Top}: Distribution of $\mu_\mathrm{LN}$ in log-scale. \textit{Bottom}: Distribution of $\sigma_\mathrm{LN}$ and its best-fit Beta-prime distribution (shown as the red curve with the parameters $\alpha=2.02$ and $\beta=8.97$).}
    \label{fig_global_LogNorm_param}
\end{figure}

To achieve this, we found the flux density distribution for all 967 CAZ light curves and parameterize them using a lognormal distribution, with the probability density function $f(x)=(x ~ \sigma \sqrt{2 \pi})^{-1} \cdot \exp \left[- ~ (\ln x - \mu)^2 / (2 \sigma ^2) \right]$. The best-fit lognormal "location" and "scale" parameters are referred to as $\mu_\mathrm{LN}$ and $\sigma_\mathrm{LN}$, respectively. For example, Fig. \ref{fig_LC_CAZJ0509+0541} shows the CAZ light curve and the median-normalized flux density distribution for the blazar J0509+0541 (TXS~0506+056) as well as its best-fit lognormal curve. Repeating this process for all light curves yields global distributions for $\mu_\mathrm{LN}$ and $\sigma_\mathrm{LN}$ as shown in Fig. \ref{fig_global_LogNorm_param}.

Since the light curves are median-normalized, we expect their $\mu_\mathrm{LN}$ to be around zero. Indeed, the observed global distribution of $\mu_\mathrm{LN}$ (top panel of Fig. \ref{fig_global_LogNorm_param}) is generally close to zero: its median is on the order of $10^{-15}$, with 90\% of values below 0.052. Notably, this distribution exhibits two peaks at $\mu_\mathrm{LN} \approx 10^{-17}$ and $\mu_\mathrm{LN} \approx 10^{-3}$, arising due to the presence of two distinct populations of flux densities in the CAZ light curves\footnote{The first half of the CAZ light curves is almost exclusively populated by CRTS light curves, while the other half contains the other surveys combined. This creates two distinct portions with different properties (e.g., cadence, signal-to-noise ratio, average flux density), often resulting in two distinct peaks in global distributions. This is explored further in the CAZ data release paper, Kouch et al. (in prep.).}. Additionally, $\mu_\mathrm{LN} \gtrsim 0.1$ in $\sim$10\% of the fits, which is likely due to low-quality light curves or poor lognormal fits. Hence, we set $\mu_\mathrm{LN}$ to zero for simplicity. On the other hand, the $\sigma_\mathrm{LN}$ global distribution shows a long-tailed Gaussian-like spread, which we fit with a Beta-prime function (see the bottom panel of Fig. \ref{fig_global_LogNorm_param}); the best-fit parameters are $\alpha=2.02$ and $\beta=8.97$. Therefore, using $\mu_\mathrm{LN}=0$ and selecting a random value from the best-fit Beta-prime distribution for $\sigma_\mathrm{LN}$, we can generate flux density distribution resembling those observed.

\section{Correlation tests} \label{sec_tests}
To search for a potential spatio-temporal correlation between blazars and neutrinos, we closely followed the methodology presented in K24, which in turn was based on our original study, H21. This correlation test begins by quantifying one or more global test-statistic (TS) parameters derived from variability measures of the spatially neutrino-associated blazars. It then determines the chance probability of these TS parameters by comparing them to those obtained from repeated Monte Carlo iterations in which the neutrinos are randomly shifted in RA\footnote{Randomizing neutrinos must only be done by shifting them in RA because neutrinos are not uniformly distributed in Dec..}. These chance probabilities are used to test and potentially reject physically motivated null-hypotheses.

In our simulations, we constructed a "spatial-only" TS parameter based on the time-averaged variability measure ($F_\mathrm{var}$) and a spatio-temporal TS parameter based on the time-resolved variability measure (AI); these are described in Sect. \ref{sec_data_blz}. In K24, we searched for a blazar--neutrino association using several test strategies, involving both spatial-only and spatio-temporal TS parameters. Since these searches were performed on real data, with the intent of establishing a correlation, all our significance levels had to be corrected for multiple trials. However, in this study, we simulated blazars with predetermined level of correlation to neutrinos (see Sect. \ref{sec_tests_sim_sample}); this allowed us to assess the detection power of each test strategy in a controlled manner. In Sects. \ref{sec_tests_TS_parameters}, \ref{sec_tests_handling_neutrinos}, and \ref{sec_tests_combining} we describe these strategies.

\subsection{Simulating the blazar samples} \label{sec_tests_sim_sample}
In this subsection we describe the simulation of the blazar samples used in this paper with varying levels of induced blazar--neutrino correlation. We then evaluate these induced correlations via various test strategies to identify the method with the highest detection power. In Sect. \ref{sec_tests_sim_sample_random} we explain the generation of a sample of 4000 random blazars with random variability measures, which serves as our initial sanity-check blazar sample. In Sect. \ref{sec_tests_sim_sample_HEN_assoc} we describe the construction of four partially neutrino-associated blazar samples with one designed to represent a realistic scenario.

\subsubsection{Random blazar sample} \label{sec_tests_sim_sample_random}
We generated a random population of 4000 blazars, each assigned with random time-averaged and time-resolved variability measures ($F_\mathrm{var}$ and AI; see Sect. \ref{sec_data_blz}). Given the uniform distribution of blazars across the extragalactic sky (e.g., \citealt{Healey2007_CRATES}), generating random sky positions is straightforward. We drew 4000 random values uniformly from $0^\circ \le \alpha < 360^\circ$ for RA ($\alpha$), and 4000 from $-90^\circ \le \delta \le 90^\circ$ for Dec. ($\delta$). Using the observed distributions of variability measures (parameterized in Sect. \ref{sec_data_blz}), we assigned a random variability measure to each blazar. A random $F_\mathrm{var}$ value was sampled from a Beta-prime distribution with parameters $\alpha=1.57$ and $\beta=5.76$, resembling the observed CAZ $F_\mathrm{var}$ distribution (see Fig. \ref{fig_global_Fvar}).

Next, we generated random AI values for each random blazar. Since AI values are associated with individual neutrino arrival time, we only required 283 AI values for each blazar. Assuming neutrino arrival times are independent\footnote{In practice this holds, since the current generation of neutrino observatories have rather poor detection rates of astrophysical neutrinos (e.g., for IceCube, this detection rate is on average $\sim$1 per month).} and that no two neutrinos from a single blazar flare are likely to be detected (e.g., \citealt{oikonomou2019_indepth_blz_neut_connection}), randomly selecting flux density values from a light curve is mathematically equivalent to randomly sampling flux density values from the flux density distribution of the light curve. Thus, for each random blazar, we drew 283 values from a random flux density distribution. These distributions were generated for each blazar using a lognormal distribution with parameters $\mu_\mathrm{LN}=0$ and $\sigma_\mathrm{LN}$ randomly sampled from a Beta-prime distribution with parameters $\alpha=2.02$ and $\beta=8.97$ (see Fig. \ref{fig_global_LogNorm_param}).

This sample constitutes our "sim-null" dataset (the purple circle in Fig. \ref{fig_sim_samples_sketch}), which is by construction purely random and serves as our initial sanity-check sample. No correlation should exist between the sim-null blazars and IceCat1+ neutrinos, although low p-values may arise due to statistical fluctuations (i.e., Gaussian statistics dictates $2\sigma$, $3\sigma$, and $4\sigma$ p-values arising in 4.55\%, 0.27\%, and $<0.01\%$ of the simulations, respectively).

\subsubsection{Partially neutrino-associated blazar samples} \label{sec_tests_sim_sample_HEN_assoc}
Next, we simulate a small set of blazars, each spatio-temporally associated with a neutrino, and add them to the sim-null sample. In total, we construct four partially neutrino-associated blazar samples: one serves as a second sanity check, two are special-case scenarios, and one represents a potentially realistic scenario.

To construct the first special-case scenario, we identify the best neutrinos in IceCat1+, specifically those with a very high probability of being astrophysical ($\mathcal{S} > 0.85$) and well-localization ($\Omega < 1~\mathrm{deg^2}$). Four neutrinos\footnote{IC101028X, IC140611A, IC171106A, and IC201007A.} meet these criteria. For each, we simulate a blazar centered on its maximum-likelihood arrival direction, with a Gaussian spread that corresponds to the $\gtrsim$90\%-likelihood error region (see Sect. \ref{sec_data_neut}). Subsequently, we assign each blazar a random $F_\mathrm{var}$ value of at least 0.37, corresponding to the top 30\% of the observed $F_\mathrm{var}$ distribution (see Fig. \ref{fig_global_Fvar}). To assign 283 random AI values to each blazar, we generate four random flux density distributions as described in Sect. \ref{sec_tests_sim_sample_random} with additional constraint that $\sigma_\mathrm{LN}$ is at least 0.1\footnote{This condition makes it easier to generate above-average AI values, which improves computation speed.}. From each distribution, we sample 282 random AI values and one AI value greater than 1.25 (see K24 for justification). Thus, each simulated blazar has a pseudo-random $F_\mathrm{var}>0.37$, 282 random AI values, and one pseudo-random $\mathrm{AI}>1.25$ (corresponding to the arrival time of the spatially associated neutrino). Adding these four pseudo-random blazars to the sim-null sample yields the "sim-best" sample, comprising 4004 blazars (the orange set in Fig. \ref{fig_blazar_neut_sketch}). This allows us to investigate the impact of four reliably reconstructed neutrino-associated blazars on the detection power of different test strategies.

To construct the second special-case blazar sample, we focus on neutrinos with $0.50 < \mathcal{S} <0.70$ and $5 < \Omega < 10~\mathrm{deg}^2$. 13 neutrinos\footnote{IC110902A, IC120515A, IC131108A, IC131124A, IC140101A, IC160225A, IC170626A, IC170704A, IC170819A, IC170923A, IC180417A, IC190730A, and IC201221A.} meet these "mid-range" criteria. Following the same process as above, we simulate 13 pseudo-random blazars with pseudo-random signals associated with these mid-range neutrinos. Adding these to sim-null sample produces the "sim-mid" sample, comprising of 4013 blazars (the green set in Fig. \ref{fig_blazar_neut_sketch}). This sample allows us to evaluate how a dozen blazars associated with mid-range neutrino events affect the detection power of various test strategies.

The remaining two simulated blazar samples represent broader scenarios. In one, all astrophysical neutrinos in IceCat1+ are assumed to be emitted by blazars ("sim-$\mathcal{S}$"), which serves as a second sanity check. The other scenario assumes that 20\% of astrophysical neutrinos are emitted by blazars ("sim-0.2$\mathcal{S}$"), representing a potentially realistic scenario. The "sim-$\mathcal{S}$" scenario is highly exaggerated as sources like NGC~1068 (\citealt{icecube_collab2022_ngc1068}) and non-AGN phenomena like tidal disruption events (\citealt{vanValzen2024_TDE_neutrinos}) and micro-quasars (\citealt{Koljonen2023_CygX3}) could be significant contributors to astrophysical neutrinos. Nevertheless, blazars are estimated to account for $\sim$20\% of astrophysical neutrinos in the $\sim$100--1000~TeV range (e.g., \citealt{Aartsen2017_blz_maximal_contr, Huber2019_blz_maximal_contr, Das2021_blazar_contribution_to_diffuse_nu, Oikonomou2022_blz_contribution_to_HEN}). Therefore, the sim-$0.2\mathcal{S}$ scenario represents a potentially realistic scenario.

In sim-$0.2\mathcal{S}$ and sim-$\mathcal{S}$, the inclusion of neutrinos is determined by their signalness ($\mathcal{S}$). For each neutrino, we drew a random number uniformly between 0 and 1. If this number exceeded $\mathcal{S}$, the neutrino was excluded from sim-$\mathcal{S}$. Similarly, if the number exceeded $0.2 \times \mathcal{S}$, the neutrino was excluded form sim-0.2$\mathcal{S}$. Repeating this for all 283 neutrinos in IceCat1+, we expected around $283 \times \widetilde{\mathcal{S}} = 283 \times 0.429 \approx 121$ neutrinos in sim-$\mathcal{S}$ and around $283 \times 0.429 \times 0.2 \approx 24$ neutrinos in sim-0.2$\mathcal{S}$. We emphasize that all $\sim$24 neutrinos in sim-0.2$\mathcal{S}$ are are also part of sim-$\mathcal{S}$, leaving $\sim$97 neutrinos unique to sim-$\mathcal{S}$.

\begin{figure}
    \centering
    \includegraphics[width=0.49\textwidth]{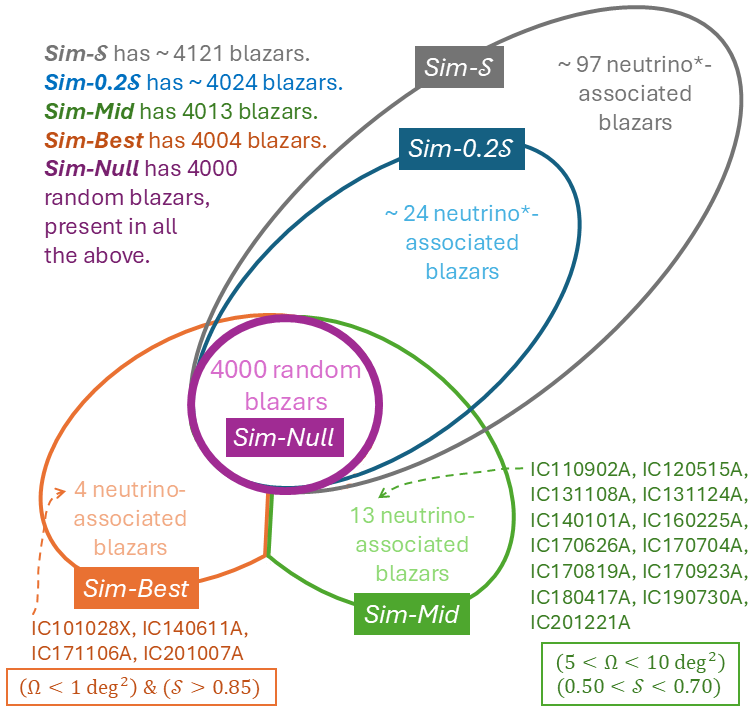}
    \caption{Venn diagram of the five simulated blazar samples. All contain the 4000 random blazars of sim-null. Sim-best and sim-mid do not overlap, whereas sim-$0.2\mathcal{S}$ is contained entirely in sim-$\mathcal{S}$. In sim-best and sim-mid, there are 4 and 13 neutrinos, respectively, which remain the same across all simulations. In sim-$\mathcal{S}$ and sim-$0.2\mathcal{S}$, neutrino* indicates their $\mathcal{S}$-based statistical nature.}
    \label{fig_sim_samples_sketch}
\end{figure}

Following the same procedure as for sim-best and sim-mid, we generated pseudo-random blazars for the accepted neutrinos in sim-$\mathcal{S}$ and sim-0.2$\mathcal{S}$. Consequently, sim-$\mathcal{S}$ and sim-0.2$\mathcal{S}$ comprise approximately 4121 and 4024 blazars, respectively, although the exact numbers vary from due to their statistical nature --- this is why we repeated generating these blazar samples 1000 times in this study. In Fig. \ref{fig_sim_samples_sketch}, sim-$\mathcal{S}$ is the gray set, and sim-0.2$\mathcal{S}$ is the blue set contained within sim-$\mathcal{S}$.

We emphasize that the 4000 random blazars of sim-null, forming the majority of each simulated sample, are identical across a single simulation loop. Therefore, any difference in the correlation strength of test strategies, between the samples in the same loop, reflects the strength of the induced signal, and not random fluctuations in sim-null. Lastly, all blazars in sim-0.2$\mathcal{S}$ are included in sim-$\mathcal{S}$. Meanwhile, apart from the 4000 random blazars, sim-best and sim-mid are mutually exclusive, though they may overlap with sim-0.2$\mathcal{S}$ and sim-$\mathcal{S}$. In Fig. \ref{fig_sim_samples_sketch} we show a Venn diagram that visualizes these relations.

\subsection{Test-statistic parameters} \label{sec_tests_TS_parameters}
When a blazar and a neutrino coincide spatially (see Sect. \ref{sec_tests_weighting}), they are included in the calculations for our correlation test. Depending on the blazar and neutrino sample sizes, there are typically hundreds of such spatial associations. Each spatial association is accompanied by a variability measure (either $F_\mathrm{var}$ or AI). To construct a single global TS parameter, we can either average a given variability measure or count the occurrences where a given variability measure exceeds a predetermined threshold value. We emphasize that these two distinct methods of calculating the global TS parameter ("averaging" and "counting") can be applied independently to each variability measure ($F_\mathrm{var}$ or AI).

The global TS parameters are calculated in both observed and Monte Carlo (random) setups (see Sect. \ref{sec_tests}). The probability ($p$) that an observed TS parameter occurs randomly is calculated as\begin{equation} \label{eqn_pval}
    p = \frac{M+1}{N+1}
,\end{equation}
where $M$ is the number of random TS parameters greater than or equal to the observed TS parameter, and $N$ is the total number of Monte Carlo iterations (\citealt{davison_hinkley_1997}). In this study, $N=10^4$, meaning the smallest detectable p-value is 0.0001 (i.e., $\sim$4$\sigma$ significance level).

When TS parameters are constructed via counting, the predetermined thresholds for $F_\mathrm{var}$ and AI are consistent with those used to simulate pseudo-random signals (see Sect. \ref{sec_tests_sim_sample_HEN_assoc}). This means that a blazar located near a neutrino event is simulated to be critically variable ($F_\mathrm{var}>0.37$) and flaring ($\mathrm{AI}>1.25$) at the arrival time of the neutrino. Therefore, the counting-based test strategies used in this study are designed to perform ideally.

\subsection{Handling poorly reconstructed neutrino events} \label{sec_tests_handling_neutrinos}
Our correlation test depends on a "spatial" coincidence between a blazar and a neutrino (see more details in Sect. \ref{sec_tests_weighting}). While the sky coordinates of blazars are typically known down to an arcsec accuracy, the sky error regions for IceCube high-energy neutrino events are generally much larger, often spanning several and sometimes tens of degrees (see Sect. \ref{sec_data_neut}). Including neutrinos with very large error regions in the analysis without proper treatment can result in spurious associations with many random blazars, reducing detection power.

Previous studies, including H21, addressed this issue by applying neutrino sample cuts, where neutrino events failing to meet predetermined criteria (e.g., error region size) were excluded from the analysis. An alternative approach is to introduce a weighting scheme, as used in K24, that reduces the contribution of poorly reconstructed events to the global TS parameter. In Sect. \ref{sec_tests_weighting} we introduce four weighting schemes that are investigated in this study. In Sect. \ref{sec_tests_neut_cuts} we explore the impact of implementing neutrino sample cuts. We note that these two techniques can be utilized simultaneously; thus, we systematically studied their individual and combined effects on the detection power of our correlation test.

\subsubsection{Weighting schemes} \label{sec_tests_weighting}
We considered four general weighting schemes in this study: two "unweighted" and two "weighted". The unweighted schemes, referred to as $\varnothing_\mathrm{W}$, act the control setups where none of the associations are weighted; for these, we used two different neutrino error region sizes (more details given below). On the other hand, the weighted schemes include a "top-hat" weighting approach ($\mathrm{W}_\mathrm{T}$) and a "Gaussian" weighting approach ($\mathrm{W}_\mathrm{G}$).

The top-hat weighting scheme, introduced in K24, uses the $\gtrsim$90\%-likelihood error region of a neutrino (see Sect. \ref{sec_data_neut}) as its spatial boundary. Here, $R_\phi$ represents the distance between the center and the edge of the error region at a given phase angle $\phi$, while $d_{\mathrm{BN},\phi}$ represents the distance between the blazar and the center of the error region along $\phi$ (see Fig. \ref{fig_blazar_neut_sketch}). The top-hat weight ($W_\mathrm{T}$) is defined as
\begin{equation} \label{eqn_W_T}
    W_\mathrm{T} =
    \begin{cases}
      0 & \text{if $d_{\mathrm{BN},\phi} > R_\phi$}\\
      \mathcal{S} & \text{if $d_{\mathrm{BN},\phi} \le R_\phi$ and $\Omega \le \widetilde{\Omega}$}\\
      \mathcal{S} \cdot \widetilde{\Omega} \ / \ \Omega & \text{if $d_{\mathrm{BN},\phi} \le R_\phi$ and $\Omega > \widetilde{\Omega}$,}
    \end{cases}
\end{equation}
where $\widetilde{\Omega}=6.63~\mathrm{deg^2}$ is the global median of $\Omega$ (see Sect. \ref{sec_data_neut}). We note that $W_\mathrm{T} \propto \Omega^{-1}$ penalizes poorly localized events when $\Omega > \widetilde{\Omega}$, while $\partial W_\mathrm{T} / \partial \Omega = 0$ ensures that well-localized events do not dominate the statistics when $\Omega \le \widetilde{\Omega}$.

If a blazar falls outside of the neutrino $\gtrsim$90\%-likelihood error region (i.e., $d_{\mathrm{BN},\phi} > R_\phi$), it is omitted from the association calculations in the top-hat weighting scheme (i.e., $W_\mathrm{T}=0$). On the other hand, $W_\mathrm{T}$ does not penalize associations based on the distance between the center of the neutrino error region and the blazar (i.e., $\partial W_\mathrm{T} / \partial d_{\mathrm{BN},\phi} = 0$). In other words, $W_\mathrm{T}$ remains the same for both on-center associations ($0 \lesssim d_{\mathrm{BN},\phi} \ll R_\phi$) and off-center ones ($d_{\mathrm{BN},\phi} \lesssim R_\phi$). Thus, if the neutrino error regions are underestimated or systematically shifted from their true centers, genuine associations could be missed. This is a potential caveat of the top-hat weighting scheme.

As an alternative, we introduced a Gaussian weighting scheme that explicitly incorporates $\mathcal{S}$, $\Omega$, and $d_{\mathrm{BN},\phi}$. It penalizes off-center spatial associations gradually, without requiring a hard "edge". The Gaussian weight ($W_\mathrm{G}$) is defined as
\begin{equation} \label{eqn_W_G}
    W_\mathrm{G} = \mathcal{S} \cdot \frac{\Omega_\mathrm{min}}{\Omega} \cdot \mathrm{exp} \left[{-\frac{1}{2} \left( \frac{2 ~ d_{\mathrm{BN},\phi}}{R_\phi} \right)^2} \right]
,\end{equation}
where $\Omega_\mathrm{min}=0.115~\mathrm{deg^2}$ (the smallest global $\Omega$) is used as a scaling factor. Figure \ref{fig_blazar_neut_sketch} visualizes the spatial parameters used in Eq. \ref{eqn_W_G}.

In the Gaussian weighting scheme all blazar--neutrino pairs are spatially "associated". However, most associations at substantial distances ($d_{\mathrm{BN},\phi} \gg R_\phi$) yield negligible $W_\mathrm{G}$; for example, $W_{\mathrm{G}}|_{d=3R}/W_{\mathrm{G}}|_{d=R}=e^{-16} \approx 10^{-7}$ implies that an association occurring at a distance of three times the reported IceCube error bar has a weight roughly seven orders of magnitude smaller than for an event occurring at the reported IceCube error bar. To optimize computational efficiency, we imposed an upper limit: if $d_{\mathrm{BN},\phi} > 3R_\phi$, then $W_\mathrm{G}=0$. The threshold of $3R_\phi$ balances precision with computational efficiency.

Thus, in this study we employed two error region thresholds ($3R_\phi$ and $R_\phi$): both are used in the unweighted schemes, while the Gaussian weighting scheme uses $3R_\phi$ and the top-hat weighting scheme uses $R_\phi$. We note that the unweighted schemes are top-hat-like, since the associations with blazars outside the error region are assigned a weight of zero, while those inside are given equal weights (i.e., a weight of one).

\subsubsection{Neutrino sample cuts} \label{sec_tests_neut_cuts}
Another method for dealing with the least reliably reconstructed neutrino events is to apply predetermined cuts to the neutrino sample. Leveraging the simulations used in this study, we can access the impact of these cuts on detection power. We used three samples of neutrinos, with one including the full set of 283 neutrinos of the IceCat1+ sample, referred to as the "no-cut" sample. A second sample, referred to as "soft-cut", is constructed by selecting all neutrinos from IceCat1+ with $\Omega<50~\mathrm{deg}^2$, resulting in 245 neutrinos. Finally, we constructed a third sample, referred to as "hard-cut", by requiring $\Omega<10~\mathrm{deg}^2$ and $\mathcal{S}>0.5$, producing a set of 71 neutrinos.

\subsection{Final correlation test strategies} \label{sec_tests_combining}
\begin{figure}
    \centering
    \includegraphics[width=0.49\textwidth]{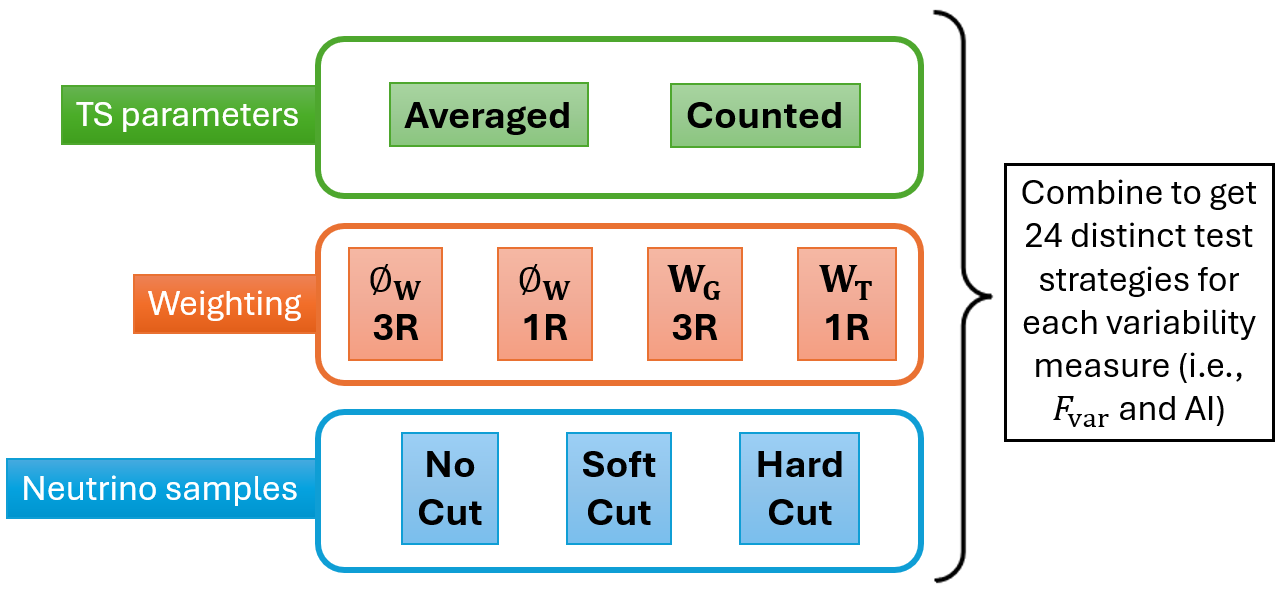}
    \caption{Visualization of the 24 possible test strategies applicable to the five blazar samples and the two variability measures. There are two options for constructing the global TS parameters (see Sect. \ref{sec_tests_TS_parameters}), four for weighting (3R and 1R refer to when $3R_\phi$ and $R_\phi$ are used as the error region edges, respectively; see Sect. \ref{sec_tests_weighting}), and three for neutrino sample cuts (see Sect. \ref{sec_tests_neut_cuts}).}
    \label{fig_correlation_tests_sketch}
\end{figure}

Figure \ref{fig_correlation_tests_sketch} visualizes our strategies for performing a correlation test on a simulated blazar sample. By combining the methods constructing the TS parameters (see Sect. \ref{sec_tests_TS_parameters}), the weighting schemes (see Sect. \ref{sec_tests_weighting}), and the different neutrino samples (see Sect. \ref{sec_tests_neut_cuts}), we derived $2 \times 4 \times 3 = 24$ unique test strategies. Since these can be applied to each of the five simulated blazar samples (see Fig. \ref{fig_sim_samples_sketch}), we were left with $24 \times 5 = 120$ correlation tests for each of the two variability measures of interest: the time-averaged $F_\mathrm{var}$ and the time-resolved AI (see Sect. \ref{sec_tests}). Thus, we obtained a total of $120 \times 2 = 240$ blazar--neutrino correlation strengths (i.e., p-values) in a single simulation step. By investigating the behavior of these p-values over 1000 simulation steps, we then determined the most optimal strategies for reliably detecting a potential blazar--neutrino correlation.

\section{Results and discussion} \label{sec_results}
Using 24 different test strategies, we obtained spatio-temporal correlation strengths (i.e., p-values) between a simulated blazar sample and IceCat1+. Since the simulated blazar sample is identical across all tests, the test strategy that results in the smallest p-value is the most optimal (i.e., it offers the greatest detection power). This comparison would not be reliable with only a single simulated blazar sample due to statistical fluctuations. Thus, we repeated the simulation process 1000 times, obtaining 1000 p-values for each test strategy to ensure robust comparisons of the p-values. Notably, since we did not perform hypotheses testing (as the underlying blazar--neutrino correlation is simulated and known), trial correction was unnecessary. The distributions of these p-values are shown in Figs. \ref{fig_pval_histograms_sim_Null_Fvar} to \ref{fig_pval_histograms_sim_Sness_AI}, where we also provide the fraction of p-values crossing the $2\sigma$, $3\sigma$, and $\sim$4$\sigma$ thresholds along with their mean and standard deviation. Here, we focus on the fraction of p-values crossing the $3\sigma$ threshold (i.e., $p<0.0027$) as a measure of quantifying and comparing the quality of different test strategies; we denote these fractions as $\mathcal{F}_{3\sigma}$. Table \ref{results_table} summarizes all $5 \times 2 \times 4 \times 3 \times 2 = 240$ values of $\mathcal{F}_{3\sigma}$ (given in \%). Below we discuss the implications of these 240 results.

\begin{table*} 
\centering 
\caption{Fraction of simulations crossing the $3\sigma$ threshold (i.e., $\mathcal{F}_{3\sigma}$ given in percent). The total number of simulations is 1000. } 
\label{results_table}
\begin{tabular}{c|c|c|rrr|rrr} 
 \hline\hline 
 & & & \multicolumn{3}{c|}{$F_\mathrm{var}$} & \multicolumn{3}{c}{AI} \\ \hline 
Sim & TS & W & No-cut & Soft-cut & Hard-cut & No-cut & Soft-cut & Hard-cut \\ 
(1) & (2) & (3) & (4) & (5) & (6) & (7) & (8) & (9) \\ 
 \hline\hline 
\multirow{8}{*}{\makecell{Sim-null\\(sanity check)}} & Averaged & $\varnothing_\mathrm{W}$ 3R & 0.1\% & 0.3\% & 0.1\% & 0.1\% & 0.1\% & 0.4\% \\ 
 & Averaged & $\varnothing_\mathrm{W}$ 1R & 0.2\% & 0.2\% & 0.2\% & 0.2\% & 0.2\% & 0.5\% \\ 
 & Counted & $\varnothing_\mathrm{W}$ 3R & 0.0\% & 0.1\% & 0.2\% & 0.1\% & 0.4\% & 0.1\% \\ 
 & Counted & $\varnothing_\mathrm{W}$ 1R & 0.2\% & 0.2\% & 0.3\% & 0.1\% & 0.3\% & 0.1\% \\ 
\cline{2-9} 
 & Averaged & $\mathrm{W}_\mathrm{G}$ 3R & 0.3\% & 0.4\% & 0.3\% & 0.1\% & 0.1\% & 0.0\% \\ 
 & Averaged & $\mathrm{W}_\mathrm{T}$ 1R & 0.2\% & 0.1\% & 0.3\% & 0.5\% & 0.4\% & 0.5\% \\ 
 & Counted & $\mathrm{W}_\mathrm{G}$ 3R & 0.3\% & 0.3\% & 0.3\% & 0.1\% & 0.1\% & 0.1\% \\ 
 & Counted & $\mathrm{W}_\mathrm{T}$ 1R  & 0.3\% & 0.3\% & 0.5\% & 0.3\% & 0.6\% & 0.1\% \\ 
\hline\hline 
\multirow{8}{*}{\makecell{Sim-best\\(special-case)}} & Averaged & $\varnothing_\mathrm{W}$ 3R & 0.2\% & 0.4\% & 0.5\% & 0.3\% & 0.1\% & 0.7\% \\ 
 & Averaged & $\varnothing_\mathrm{W}$ 1R & 0.2\% & 0.6\% & 0.6\% & 0.4\% & 0.2\% & 0.6\% \\ 
 & Counted & $\varnothing_\mathrm{W}$ 3R & 0.0\% & 0.4\% & 2.2\% & 0.4\% & 1.1\% & 4.7\% \\ 
 & Counted & $\varnothing_\mathrm{W}$ 1R & 0.3\% & 0.5\% & 4.7\% & 0.4\% & 1.2\% & 8.8\% \\ 
\cline{2-9} 
 & Averaged & $\mathrm{W}_\mathrm{G}$ 3R & 89.6\% & 89.1\% & 84.4\% & 83.6\% & 83.8\% & 83.7\% \\ 
 & Averaged & $\mathrm{W}_\mathrm{T}$ 1R & 5.9\% & 6.3\% & 2.8\% & 2.2\% & 2.3\% & 2.0\% \\ 
 & Counted & $\mathrm{W}_\mathrm{G}$ 3R & 96.7\% & 96.7\% & 96.8\% & 98.7\% & 98.7\% & 98.7\% \\ 
 & Counted & $\mathrm{W}_\mathrm{T}$ 1R  & 8.4\% & 9.3\% & 20.1\% & 24.0\% & 25.3\% & 47.4\% \\ 
\hline\hline 
\multirow{8}{*}{\makecell{Sim-mid\\(special-case)}} & Averaged & $\varnothing_\mathrm{W}$ 3R & 0.5\% & 1.7\% & 3.0\% & 0.3\% & 0.2\% & 1.2\% \\ 
 & Averaged & $\varnothing_\mathrm{W}$ 1R & 0.9\% & 1.8\% & 1.0\% & 0.7\% & 0.5\% & 0.4\% \\ 
 & Counted & $\varnothing_\mathrm{W}$ 3R & 0.8\% & 2.7\% & 55.7\% & 1.5\% & 9.9\% & 97.2\% \\ 
 & Counted & $\varnothing_\mathrm{W}$ 1R & 2.4\% & 8.8\% & 94.4\% & 4.4\% & 25.3\% & 100.0\% \\ 
\cline{2-9} 
 & Averaged & $\mathrm{W}_\mathrm{G}$ 3R & 0.4\% & 0.4\% & 0.1\% & 0.2\% & 0.2\% & 0.0\% \\ 
 & Averaged & $\mathrm{W}_\mathrm{T}$ 1R & 25.2\% & 17.8\% & 0.2\% & 8.9\% & 5.4\% & 0.1\% \\ 
 & Counted & $\mathrm{W}_\mathrm{G}$ 3R & 0.8\% & 0.8\% & 0.8\% & 0.4\% & 0.4\% & 0.2\% \\ 
 & Counted & $\mathrm{W}_\mathrm{T}$ 1R  & 46.8\% & 49.2\% & 82.3\% & 87.5\% & 87.9\% & 99.7\% \\ 
\hline\hline 
\multirow{8}{*}{\makecell{Sim-$0.2\mathcal{S}$\\(realistic)}} & Averaged & $\varnothing_\mathrm{W}$ 3R & 1.8\% & 4.2\% & 1.7\% & 0.6\% & 0.6\% & 0.6\% \\ 
 & Averaged & $\varnothing_\mathrm{W}$ 1R & 3.9\% & 7.4\% & 0.6\% & 0.5\% & 0.7\% & 0.4\% \\ 
 & Counted & $\varnothing_\mathrm{W}$ 3R & 5.0\% & 14.2\% & 24.8\% & 16.7\% & 44.2\% & 55.6\% \\ 
 & Counted & $\varnothing_\mathrm{W}$ 1R & 15.6\% & 43.2\% & 55.1\% & 36.9\% & 81.7\% & 80.3\% \\ 
\cline{2-9} 
 & Averaged & $\mathrm{W}_\mathrm{G}$ 3R & 53.6\% & 50.0\% & 31.8\% & 46.8\% & 42.9\% & 27.9\% \\ 
 & Averaged & $\mathrm{W}_\mathrm{T}$ 1R & 58.9\% & 41.7\% & 2.1\% & 26.6\% & 14.8\% & 1.0\% \\ 
 & Counted & $\mathrm{W}_\mathrm{G}$ 3R & 68.7\% & 68.6\% & 58.8\% & 83.6\% & 83.0\% & 72.9\% \\ 
 & Counted & $\mathrm{W}_\mathrm{T}$ 1R  & 84.8\% & 84.8\% & 69.7\% & 97.8\% & 97.5\% & 88.4\% \\ 
\hline\hline 
\multirow{8}{*}{\makecell{Sim-$\mathcal{S}$\\(sanity check)}} & Averaged & $\varnothing_\mathrm{W}$ 3R & 61.7\% & 80.0\% & 38.9\% & 14.6\% & 25.0\% & 13.0\% \\ 
 & Averaged & $\varnothing_\mathrm{W}$ 1R & 88.2\% & 95.3\% & 2.2\% & 29.0\% & 29.3\% & 1.3\% \\ 
 & Counted & $\varnothing_\mathrm{W}$ 3R & 99.6\% & 100.0\% & 100.0\% & 100.0\% & 100.0\% & 100.0\% \\ 
 & Counted & $\varnothing_\mathrm{W}$ 1R & 100.0\% & 100.0\% & 100.0\% & 100.0\% & 100.0\% & 100.0\% \\ 
\cline{2-9} 
 & Averaged & $\mathrm{W}_\mathrm{G}$ 3R & 100.0\% & 100.0\% & 98.8\% & 97.5\% & 97.8\% & 95.5\% \\ 
 & Averaged & $\mathrm{W}_\mathrm{T}$ 1R & 100.0\% & 100.0\% & 13.5\% & 88.9\% & 77.7\% & 7.1\% \\ 
 & Counted & $\mathrm{W}_\mathrm{G}$ 3R & 100.0\% & 100.0\% & 100.0\% & 100.0\% & 100.0\% & 100.0\% \\ 
 & Counted & $\mathrm{W}_\mathrm{T}$ 1R  & 100.0\% & 100.0\% & 100.0\% & 100.0\% & 100.0\% & 100.0\% \\ 
\hline\hline 
\end{tabular}
\tablefoot{Column (1) refers to the sample of simulated blazars (see Sect. \ref{sec_tests_sim_sample}). Column (2) refers to the TS parameter (see Sect. \ref{sec_tests_TS_parameters}); Averaged and Counted refer to the averaging-based and counting-based TS parameters, respectively. Column (3) refers to the weighting scheme as well as the neutrino error region size (see Sect. \ref{sec_tests_weighting}); $\varnothing_\mathrm{W}$, $\mathrm{W}_\mathrm{G}$, and $\mathrm{W}_\mathrm{T}$ imply no weighting, Gaussian weighting, and top-hat weighting, respectively; 1R and 3R indicate the neutrino error region edge is $R_\phi$ and $3R_\phi$, respectively. Columns (4) and (7) refer to the no-cut neutrino subsample, columns (5) and (8) refer to the soft-cut neutrino subsample, and columns (6) and (9) refer to the hard-cut neutrino subsample (see Sect. \ref{sec_tests_neut_cuts}). In columns (4)--(6), $F_\mathrm{var}$ is used as the variability measure; while in columns (7)--(9), AI is used as the variability measure.}
\end{table*}

\subsection{Sanity checks: sim-null and sim-$\mathcal{S}$} \label{sec_results_sanity_checks}
The sim-null blazar sample does not have any simulated blazar-neutrino connection (see Sect. \ref{sec_tests_sim_sample_random}); whereas, in sim-$\mathcal{S}$, all astrophysical neutrinos are simulated as originating from blazars (see Sect. \ref{sec_tests_sim_sample_HEN_assoc}). These samples serve as sanity checks of our simulation setup and correlation test results. As shown in Table \ref{results_table}, under all test strategies, the sim-null results are consistent with random fluctuations occurring within the $3\sigma$ limits; the average $\mathcal{F}_{3\sigma}$ is 0.23\% (with a standard deviation of 0.14\%), which closely matches the chance probability of obtaining a $3\sigma$ result from a completely random process (i.e., $p \approx 0.0027$).

In sim-$\mathcal{S}$, most $\mathcal{F}_{3\sigma}$ are 100\% (the average $\mathcal{F}_{3\sigma}$ is 80.3\%, with a few test strategies yielding notably lower fractions). This demonstrates two key points: (1) most test strategies can consistently detect $3\sigma$ (even $\sim$4$\sigma$; see Appendix \ref{appendix_pval_dist}) blazar--neutrino associations; and (2) certain test strategies clearly underperform even when detecting the same underlying blazar--neutrino correlation. For example, combining the hard-cut sample with averaging in the top-hat weighting scenario results in poor detection power.

\subsection{Realistic scenario: sim-0.2$\mathcal{S}$} \label{sec_results_realistic}
In sim-$0.2\mathcal{S}$, 20\% of astrophysical neutrinos are simulated to be associated with blazars, a ratio compatible with observations (see Sect. \ref{sec_tests_sim_sample_HEN_assoc}). As shown in Table \ref{results_table}, $\mathcal{F}_{3\sigma}$ ranges from $\sim$98\% to $\sim$0\%, which allows us to systematically disfavor underperforming strategies. Crucially, changing from sim-$\mathcal{S}$ to sim-$0.2\mathcal{S}$ results in reduced $\mathcal{F}_{3\sigma}$ for all test strategies, as expected.

\subsubsection{Effect of averaging versus counting} \label{sec_results_realistic_avg_vs_cnt}
We first compared the average $\mathcal{F}_{3\sigma}$ for tests whose global TS parameter was constructed via averaging to those  constructed via counting in the sim-$0.2\mathcal{S}$ scenario. The average $\mathcal{F}_{3\sigma}$ is 17.5\% in the former and 59.7\% in the latter. This clearly shows that counting consistently outperforms averaging. However, we note that the thresholds used in the counting strategies are ideal (see Sect. \ref{sec_tests_TS_parameters}).

\subsubsection{Effect of weighting} \label{sec_results_realistic_weighting}
$\mathcal{F}_{3\sigma}$ increases when changing from unweighted ($\varnothing_\mathrm{W}$ 3R and $\varnothing_\mathrm{W}$ 1R) test strategies to their respective weighted ones ($\mathrm{W}_\mathrm{G}$ 3R and $\mathrm{W}_\mathrm{T}$ 1R). When considering only the counted results in sim-$0.2\mathcal{S}$, the average $\mathcal{F}_{3\sigma}$ for unweighted tests is 39.4\% compared to 79.9\% for the weighted ones. This demonstrates that utilizing a weighting scheme universally improves detection power, regardless of the test strategy. Further differences between Gaussian and top-hat weighting schemes are explored in Sect. \ref{sec_results_realistic_Gaussian_v_TopHat}.

\subsubsection{Effect of setting neutrino sample cuts} \label{sec_results_realistic_cuts}
For weighted and counted tests in sim-$0.2\mathcal{S}$, $\mathcal{F}_{3\sigma}$ decreases when progressing from the no-cut to soft-cut and then to hard-cut tests (average $\mathcal{F}_{3\sigma}$ is 83.7\%, 83.5\%, and 72.4\%, respectively). This indicates that once weighting is applied, removing neutrinos from the analysis generally reduces detection power. As argued in K24, weighting is preferable for suppressing noise from poorly reconstructed neutrinos. Another caveat of setting hard cuts is increased likelihood of false-negatives when the TS parameter is expected to be a small, whole number (e.g., in counted-AI tests; see Appendix \ref{appendix_pval_dist}). Therefore, we recommend using the entire sample of neutrinos in combination with a weighting scheme. However, if sample cuts are necessary for practical reasons (e.g., increasing computational efficiency), they should be kept as minimal as possible. 

\subsubsection{Weighting scheme: Gaussian versus top-hat} \label{sec_results_realistic_Gaussian_v_TopHat}
In the optimal test strategy where counting and weighting (with no cuts) are employed, the average $\mathcal{F}_{3\sigma}$ in the Gaussian ($\mathrm{W}_\mathrm{G}$ 3R) and top-hat ($\mathrm{W}_\mathrm{T}$ 1R) weighting schemes are 76.2\% and 91.3\%, respectively. This indicates that the top-hat weighting scheme offers greater detection power than the Gaussian one. Despite the simulated errors being Gaussian, the superior performance of the top-hat scheme can be understood by examining the results of the special-case tests, which are discussed below.

\subsubsubsection{Special-case tests: sim-best and sim-mid} \label{sec_results_realistic_Gaussian_v_TopHat_special_cases}
In sim-best, the simulated signal originates from four of the most reliably reconstructed neutrinos; whereas in sim-mid, it comes from 13 mid-range ones (see Sect. \ref{sec_tests_sim_sample_HEN_assoc}). Using Gaussian weighting (in the no-cut and counted test), $\mathcal{F}_{3\sigma}$ for sim-best is 96.7\%, while for sim-mid, the same strategy results in $\mathcal{F}_{3\sigma}=0.8\%$. However, with top-hat weighting, sim-best gives $\mathcal{F}_{3\sigma}=8.4\%$, while sim-mid gives $\mathcal{F}_{3\sigma}=46.8\%$. Strikingly, in sim-best, Gaussian weighting outperforms even the sim-$0.2\mathcal{S}$, which demonstrates its extreme sensitivity to the most reliably reconstructed events (i.e., only a few well-localized associations can dominate its statistics). This hampers the usefulness of Gaussian weighting in population-based blazar--neutrino correlation tests, because mid-range events --- even after accounting for their lower signalness --- are more numerous than well-localized neutrinos. For example, there are 220 neutrinos satisfying $0.2<\mathcal{S}<0.6$ compared to 56 satisfying $0.6<\mathcal{S}<1.0$; after adjusting for signalness, we expected $220 \times 0.4 = 88$ and $56 \times 0.8 \approx 45$ astrophysical neutrinos in each category to be potentially associated with blazars. While many mid-range neutrinos have rather poor sky localizations (i.e., large $\Omega$), they contain the majority of potential blazar--neutrino associations. Therefore, the ability of top-hat weighting to recover signal from these mid-range events allows it to outperform Gaussian weighting on average. Thus, we conclude that the top-hat weighting scheme is generally preferable in the context of spatio-temporal correlation tests.

\subsubsubsection{Robustness of the top-hat weighting scheme} \label{sec_results_realistic_Gaussian_v_TopHat_robustness_of_TH}
A potential disadvantage of the top-hat weighting (as compared to Gaussian) is its dependence on the choice of error region size (see Sect. \ref{sec_tests_weighting}). However, here we demonstrate its robustness against this parameter. We compared the results of the unweighted ($\varnothing_\mathrm{W}$) 1R and 3R tests, whose contributions to the statistics are top-hat-like (see Sect. \ref{sec_tests_weighting}). We note that: (1) this 1R-to-3R comparison is not possible for $\mathrm{W}_\mathrm{T}$ as it does not have a 3R counterpart; and (2) this comparison is not applicable to $\mathrm{W}_\mathrm{G}$ by definition. When changing from $\varnothing_\mathrm{W}$ 1R to 3R in sim-null, increasing the neutrino error region size does not arbitrarily induce false-positives. Meanwhile, when a real blazar--neutrino association exists (e.g., sim-$0.2\mathcal{S}$), some unweighted 3R tests result in $3\sigma$ associations. Interestingly, the detection power in the unweighted tests tends to decrease when changing from 1R to 3R, implying that enlarging the error region provides conservative estimate of the correlation strength. In summary, using $\varnothing_\mathrm{W}$ as a tracer of top-hat-like weighting schemes shows that: (1) the choice of the error region size does not result in false-positives; and (2) enlarging the error region size yields conservative estimates of the strength of the underlying blazar--neutrino correlation. The latter point is especially useful for correlation tests using IceCube high-energy neutrino events, as these may include unknown systematic errors\footnote{This can be due to various factors, which may include: the unknown kinematic angle between the neutrino and the muon, the unknown scattering properties of the ice, and the lack of knowledge about the exact location of the optical modules.} (e.g., \citealt{aartsen2013_syserr_1deg, abbasi2021_syserr_related}). Consequently, several studies have enlarged the IceCube error regions when searching for blazar--neutrino correlations (e.g., \citealt{plavin2020}, H21, K24), the robustness of which is now confirmed by our simulations.

\subsubsection{Variability measure: $F_\mathrm{var}$ versus AI} \label{sec_results_realistic_Fvar_vs_AI}
Finally, we examined the results of the two variability measures, $F_\mathrm{var}$ and AI. As discussed in Sect. \ref{sec_intro}, these measures test different physical hypotheses regarding the connection between blazar variability and neutrino emission. When neutrinos are produced during individual flares (as opposed to being produced in highly variable sources), the number of $3\sigma$ associations is on average higher. Nevertheless, we emphasize that testing for both hypotheses using the optimal test strategies is feasible, as both yield relatively high fractions of $3\sigma$ associations.

\subsection{Implications for H21 and K24} \label{sec_results_implications_for_H21_and_K24}
With the identification of counting and top-hat weighting (without neutrino sample cuts) as the most optimal strategy, we now revisit our previous studies (H21 and K24), where we used flux density and AI to construct the TS parameters (in the radio band for H21, and in the radio and optical bands for K24). While we cannot directly comment on average flux density as it is not a variability measure in this study, we can provide insights into AI, although we note that its definition somewhat differed in those studies (see Sect. \ref{sec_data_blz}). In both H21 and K24, we considered one averaging and two counting scenarios to construct the global AI-based TS parameter. In H21, we employed a cut on the neutrino sample comparable to the hard-cut used in this study (no weighting was utilized). Conversely, in K24, we omitted neutrino sample cuts but introduced and employed top-hat weighting. First, in both studies, the use of counting (rather than averaging) resulted in the strongest associations, aligning with our findings here when a real blazar--neutrino association exists. Second, in hindsight, the search criteria employed in K24 appear to have been the most optimal and resulted in $p=0.0377$ (before trial correction\footnote{In K24, we did not obtain individual post-trial p-values, only a global post-trial p-value, which is not directly comparable to this test strategy.}, under the most directly comparable scenario). This p-value is significantly higher than expected if 20\% of astrophysical neutrinos are associated with blazars (in which case our simulations predict a 97.8\% chance of $p<0.0027$). This suggests that the fraction of astrophysical neutrinos from blazars is substantially lower than 20\%. However, we strongly caution against over-interpreting this implication. In K24, the analysis was based on 1061 optical light curves whereas in this study we use $\sim$4000 light curves, and as shown in L22, the number of light curves is critically important to the robustness of the results.

\section{Summary and conclusions} \label{sec_conclusions}
In this study we simulated five different blazar samples with varying degrees of spatio-temporal correlation to high-energy neutrino events from IceCat1+ (see Sects. \ref{sec_data_neut},  \ref{sec_data_blz}, and \ref{sec_tests_sim_sample}). Two of these blazar samples served as sanity checks, two address special-case scenarios, and one represents a potentially realistic test. We then investigated the detection power of our spatio-temporal correlation test using 24 different strategies (see Sect. \ref{sec_tests_combining}). Namely, we considered the impact of constructing TS parameters using time-averaged and time-resolved variability measures, employing both averaging-based and counting-based methods (see Sect. \ref{sec_tests_TS_parameters}). We also investigated the effects of using unweighted or weighted schemes (top-hat and Gaussian; see Sect. \ref{sec_tests_weighting}) and considered the implications of applying cuts to the neutrino sample (see Sect. \ref{sec_tests_neut_cuts}). The results and their implications are discussed in Sect. \ref{sec_results}. Below, we summarize the most important findings:
\begin{itemize}
    \item Both time-averaged and time-resolved variability measures can achieve $\sim$4$\sigma$ blazar--neutrino associations.
    \item When calculating the global TS parameters, counting-based methods outperform averaging-based approaches.
    \item Weighing down the associations involving poorly reconstructed events is more effective than applying arbitrary cuts to the neutrino sample.
    \item Top-hat weighting is preferable to Gaussian weighting, as it is less sensitive to the most reliably reconstructed events and better recovers signals from the more numerous mid-range events.
    \item When utilizing top-hat-like weighting schemes, enlarging the neutrino error region does not arbitrarily increase the number of false-positives, and such enlargements can help in estimating the underlying correlation conservatively.
\end{itemize}

We highlight two caveats associated with this study: (1) while we consider signalness to ideally represent the likelihood that an IceCube high-energy neutrino event is of astrophysical origin, in practice, it is only a rough estimate; and (2) the counting scheme is constructed to be optimal (i.e., the neutrino-associated blazars are considered highly variable), and thus the conclusion that the counting-based approach is universally better may not hold as strongly in practice. Nevertheless, these simulations provide a potential method for further constraining the maximum fraction of astrophysical neutrinos originating from blazars.

In general, this kind of a priori fine-tuning of test strategies is essential for addressing the blazar--neutrino connection without introducing unnecessary trials when analyzing real data, especially with the advent of next-generation scientific instruments. For example, the presented methods are particularly useful for determining whether specific subpopulations of blazars are stronger neutrino emitters (e.g., certain high synchrotron peaked blazars, as suggested by \citealt{padovani2016_HSPs_as_neut_emitters}).

The next generation of neutrino observatories such as IceCube-Gen2 \citep{aartsen2014_icecube_gen2}, KM3NeT \citep{k3mnet_collab2016_next_gen}, Baikal-GVD \citep{shoibonov2012_bikal_gvd_upgrade}, RNO-G \citep{aguilar2021_greenland_neut_obs}, and GRAND \citep{fang2017_china_neut_obs} are under development. Additionally, upcoming all-sky surveys, such as the \textit{Vera C. Rubin} Observatory’s Legacy Survey of Space and Time (LSST; \citealt{LSST2009}), the Cosmic Microwave Background Stage~4 project (CMB-S4; \citealt{Abazajian2022_CMB_S4}), and the Simons Observatory (SO; \citealt{Ade2019_Simons_Observatory}), will revolutionize transient astronomy and provide an unprecedented number of light curves with unparalleled cadence for variable sources (several of which are potential neutrino factories, e.g., blazars, AGNs in general,  tidal disruption events, etc.). Moreover, there are already plans to study the central engines of neutrino-emitting blazars down to (sub)parsec-scale resolutions with the next generation Event Horizon Telescope (ngEHT; e.g., \citealt{Lico2023_ngEHT, Kovalev2023_future_of_neutrinos_w_VLBI}). Therefore, studies similar to those presented here are critical for developing the optimal test strategies needed to effectively analyze the wealth of data from these future observatories.

\begin{acknowledgements}
      We thank the anonymous referee for their highly insightful and constructive comments that helped improve and clarify the paper.
      P.K. was supported by Academy of Finland projects 346071 and 345899.
      E.L. was supported by Academy of Finland projects 317636, 320045, and 346071.
      T.H. was supported by Academy of Finland projects 317383, 320085, 322535, and 345899.
      J.J. was supported by Academy of Finland projects 320085 and 345899. 
      K.K. acknowledges support from the European Research Council (ERC) under the European Union’s Horizon 2020 research and innovation programme (grant agreement No. 101002352).      
      Based on observations obtained with the Samuel Oschin Telescope 48-inch and the 60-inch Telescope at the Palomar Observatory as part of the \textit{Zwicky} Transient Facility project. ZTF is supported by the National Science Foundation under Grant No. AST-2034437 and a collaboration including Caltech, IPAC, the Weizmann Institute for Science, the Oskar Klein Center at Stockholm University, the University of Maryland, Deutsches Elektronen-Synchrotron and Humboldt University, the TANGO Consortium of Taiwan, the University of Wisconsin at Milwaukee, Trinity College Dublin, Lawrence Livermore National Laboratories, and IN2P3, France. Operations are conducted by COO, IPAC, and UW. The ZTF forced-photometry service was funded under the Heising-Simons Foundation grant \#12540303 (PI: M.J.Graham).      
      This work has made use of data from the Asteroid Terrestrial-impact Last Alert System (ATLAS) project. The Asteroid Terrestrial-impact Last Alert System (ATLAS) project is primarily funded to search for near earth asteroids through NASA grants NN12AR55G, 80NSSC18K0284, and 80NSSC18K1575; byproducts of the NEO search include images and catalogs from the survey area. This work was partially funded by Kepler/K2 grant J1944/80NSSC19K0112 and HST GO-15889, and STFC grants ST/T000198/1 and ST/S006109/1. The ATLAS science products have been made possible through the contributions of the University of Hawaii Institute for Astronomy, the Queen’s University Belfast, the Space Telescope Science Institute, the South African Astronomical Observatory, and The Millennium Institute of Astrophysics (MAS), Chile.
      This work has made use of data from the Joan Oró Telescope (TJO) of the Montsec Observatory (OdM), which is owned by the Catalan Government and operated by the Institute for Space Studies of Catalonia (IEEC).
\end{acknowledgements}

\bibliographystyle{aa} 
\bibliography{ref.bib} 

\begin{appendix}

\section{Distributions of the simulation p-values} \label{appendix_pval_dist}
In this appendix we provide the p-value distributions of the 240 different correlation tests studied in this paper. Each of the 240 distributions contains 1000 p-values corresponding to 1000 simulation steps. Within each simulation step, we obtained a p-value by running a spatio-temporal correlation test on a sample of high-energy neutrino events and blazars (see Sect. \ref{sec_tests}).

\begin{figure*}
    \centering
    \includegraphics[width=17cm, keepaspectratio]{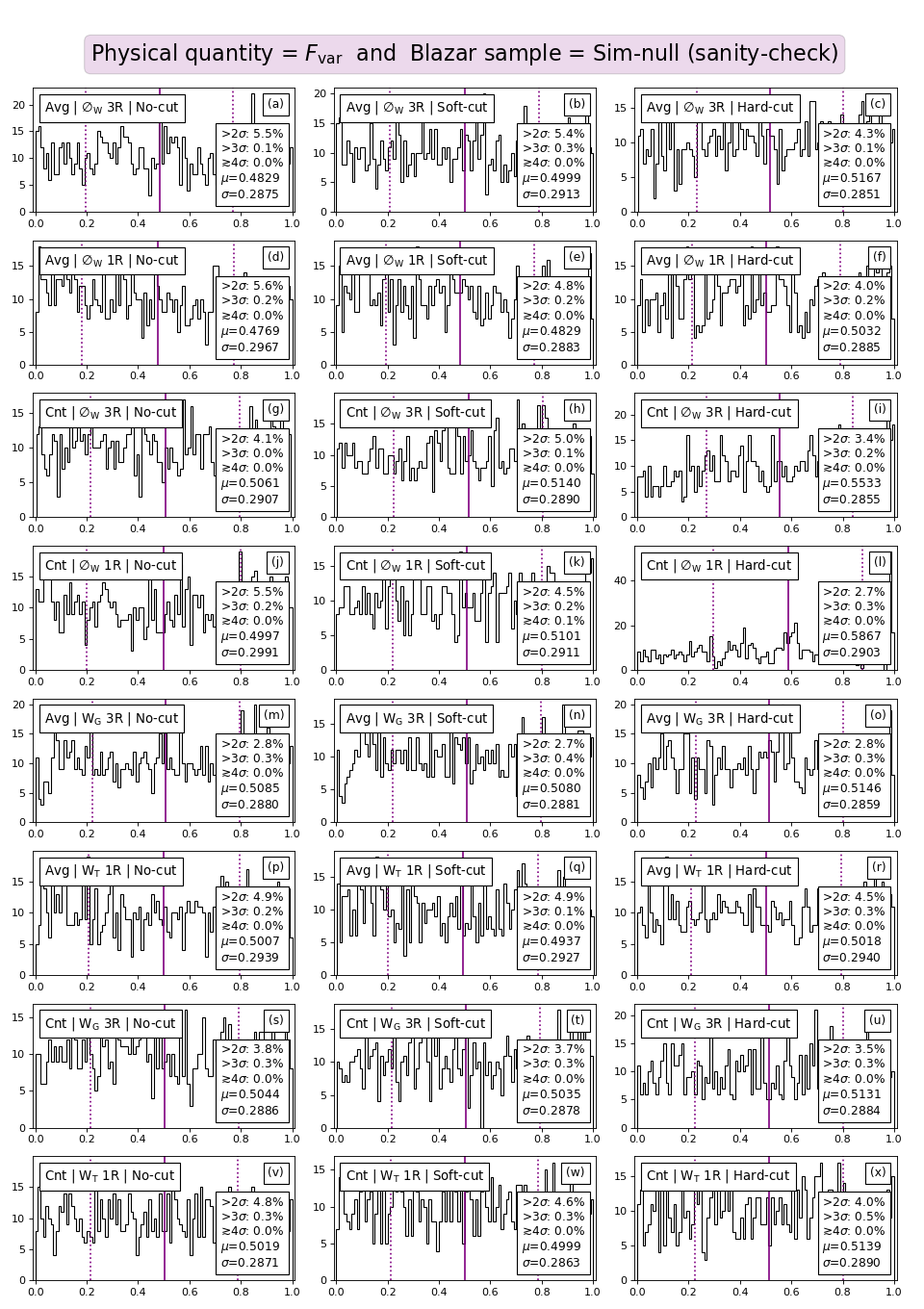}
    \caption{P-value distributions when using the time-averaged $F_\mathrm{var}$ and the sim-null blazars. For more details, see the text in Appendix \ref{appendix_pval_dist}.}
    \label{fig_pval_histograms_sim_Null_Fvar}
\end{figure*}

\begin{figure*}
    \centering
    \includegraphics[width=17cm, keepaspectratio]{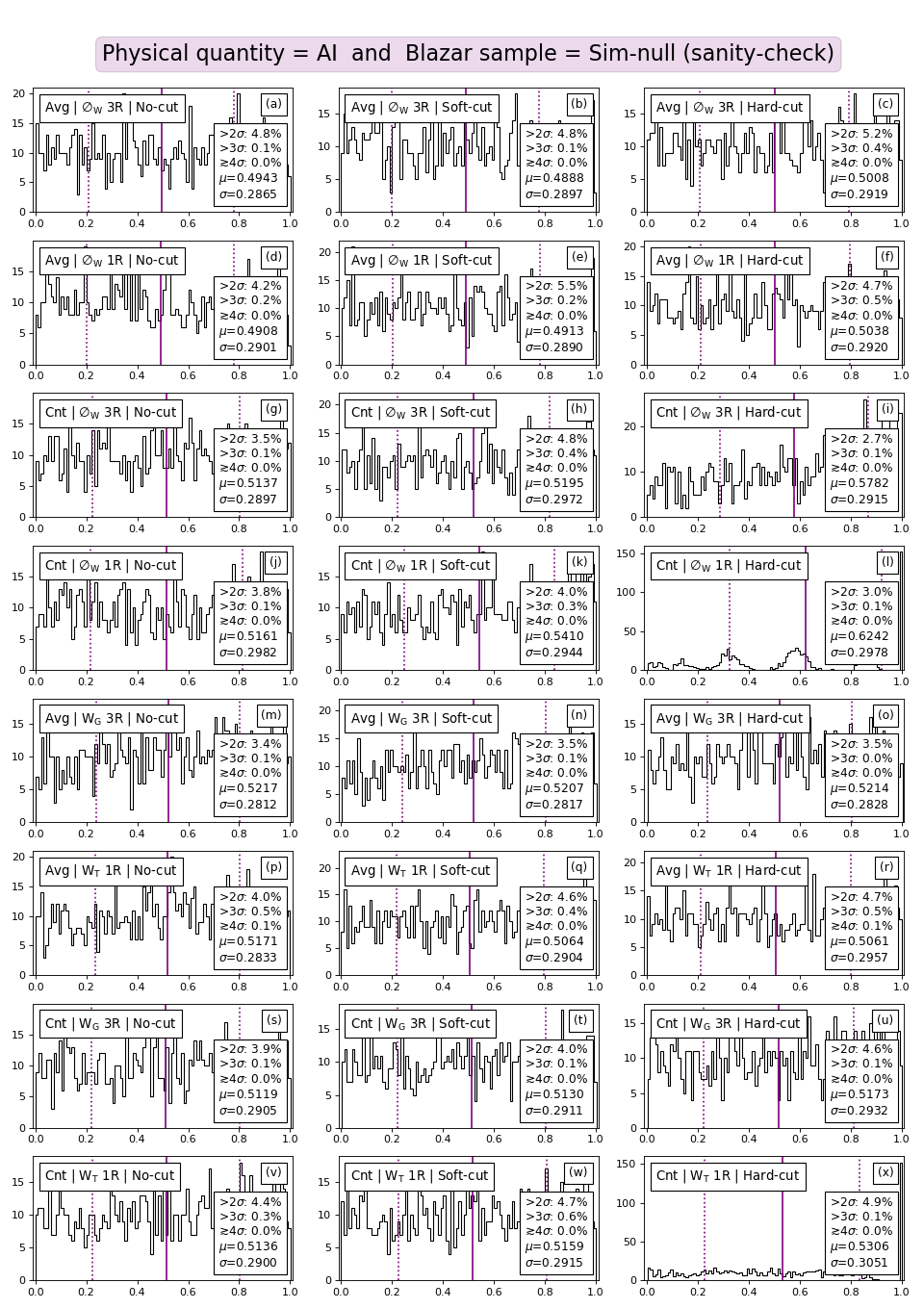}
    \caption{P-value distributions when using the time-resolved AI and the sim-null blazars. For more details, see the text in Appendix \ref{appendix_pval_dist}.}
    \label{fig_pval_histograms_sim_Null_AI}
\end{figure*}

\begin{figure*}
    \centering
    \includegraphics[width=17cm, keepaspectratio]{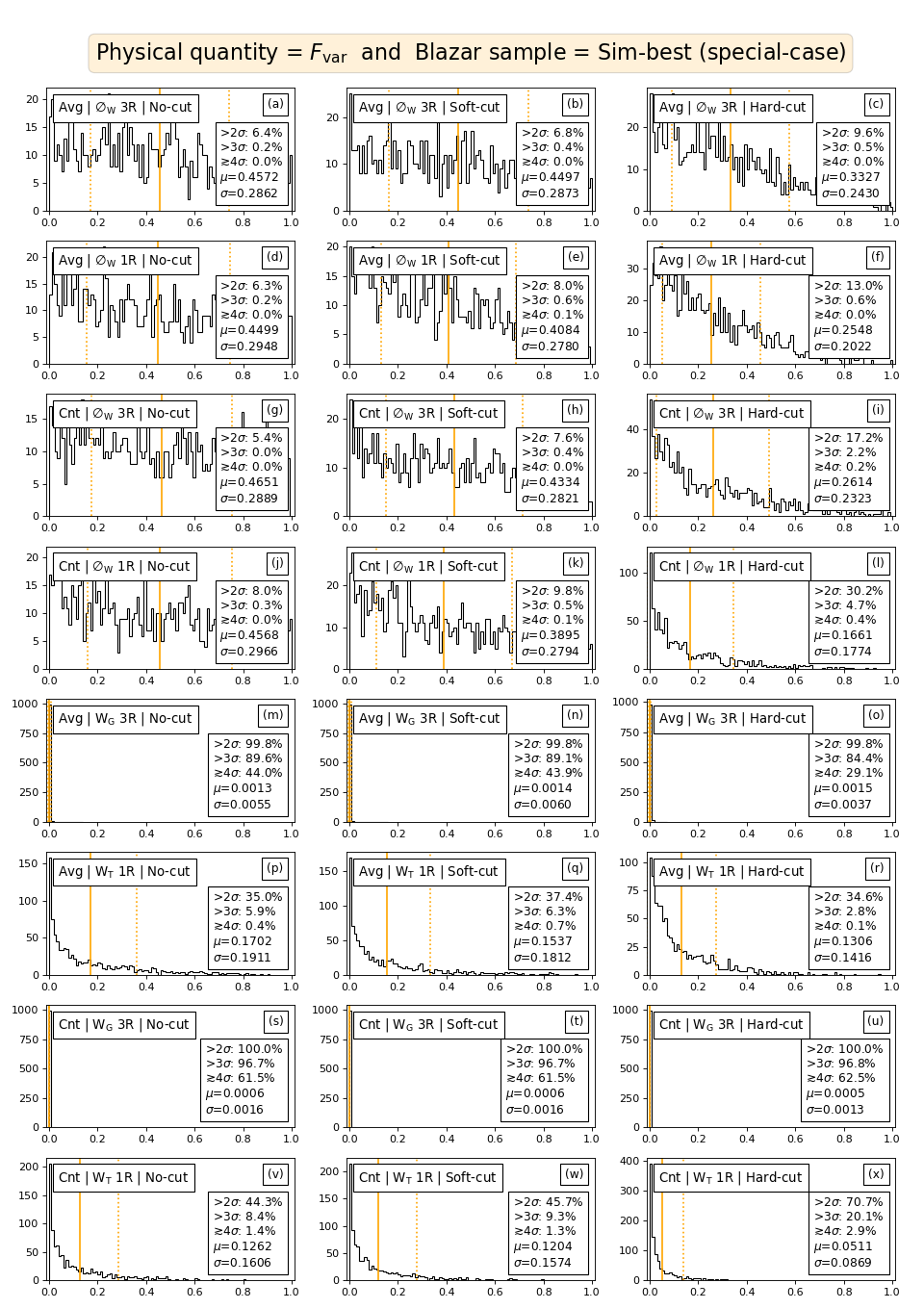}
    \caption{P-value distributions when using the time-averaged $F_\mathrm{var}$ and the sim-best blazars. For more details, see the text in Appendix \ref{appendix_pval_dist}.}
    \label{fig_pval_histograms_sim_Best_Fvar}
\end{figure*}

\begin{figure*}
    \centering
    \includegraphics[width=17cm, keepaspectratio]{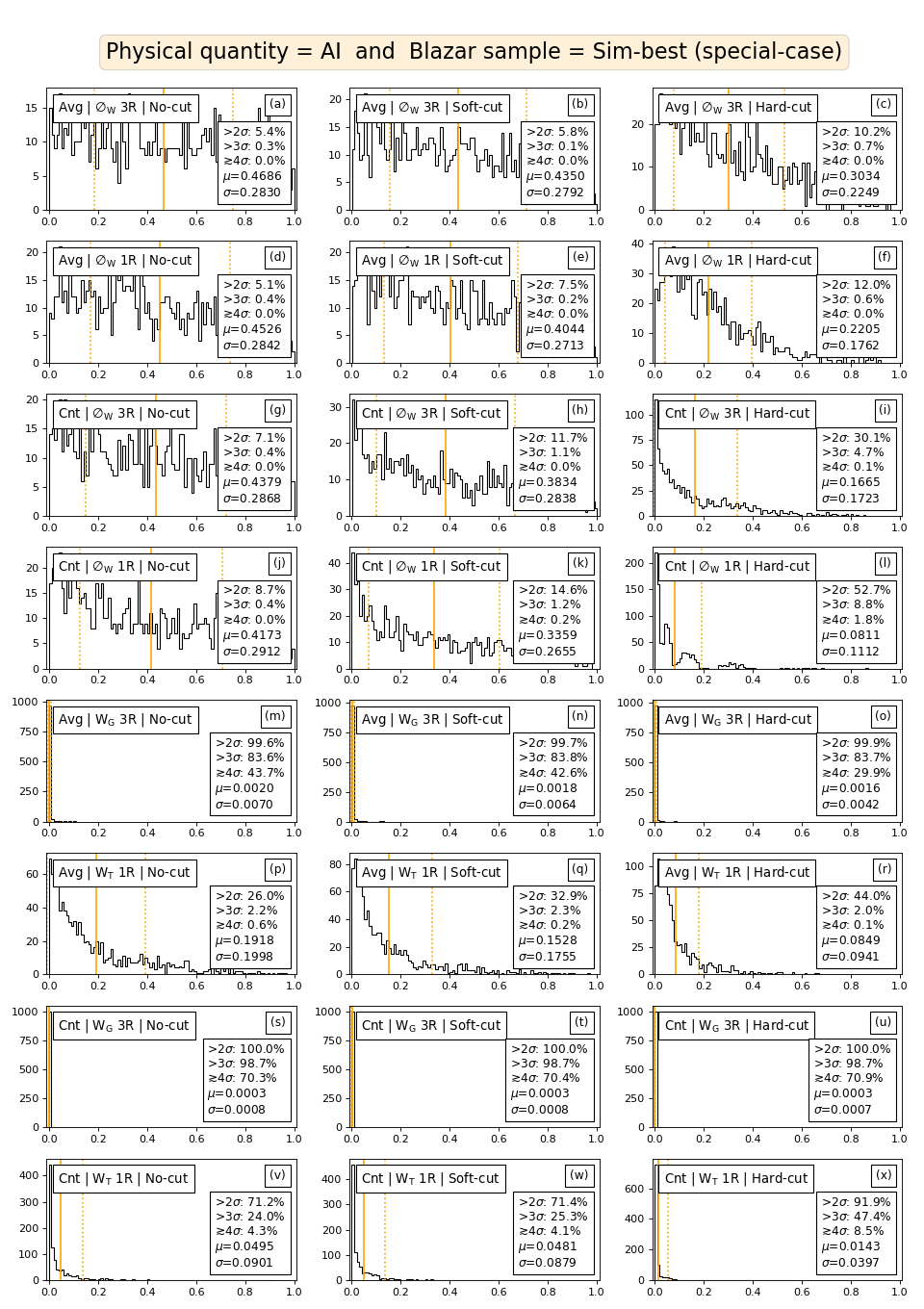}
    \caption{P-value distributions when using the time-resolved AI and the sim-best blazars. For more details, see the text in Appendix \ref{appendix_pval_dist}.}
    \label{fig_pval_histograms_sim_Best_AI}
\end{figure*}

\begin{figure*}
    \centering
    \includegraphics[width=17cm, keepaspectratio]{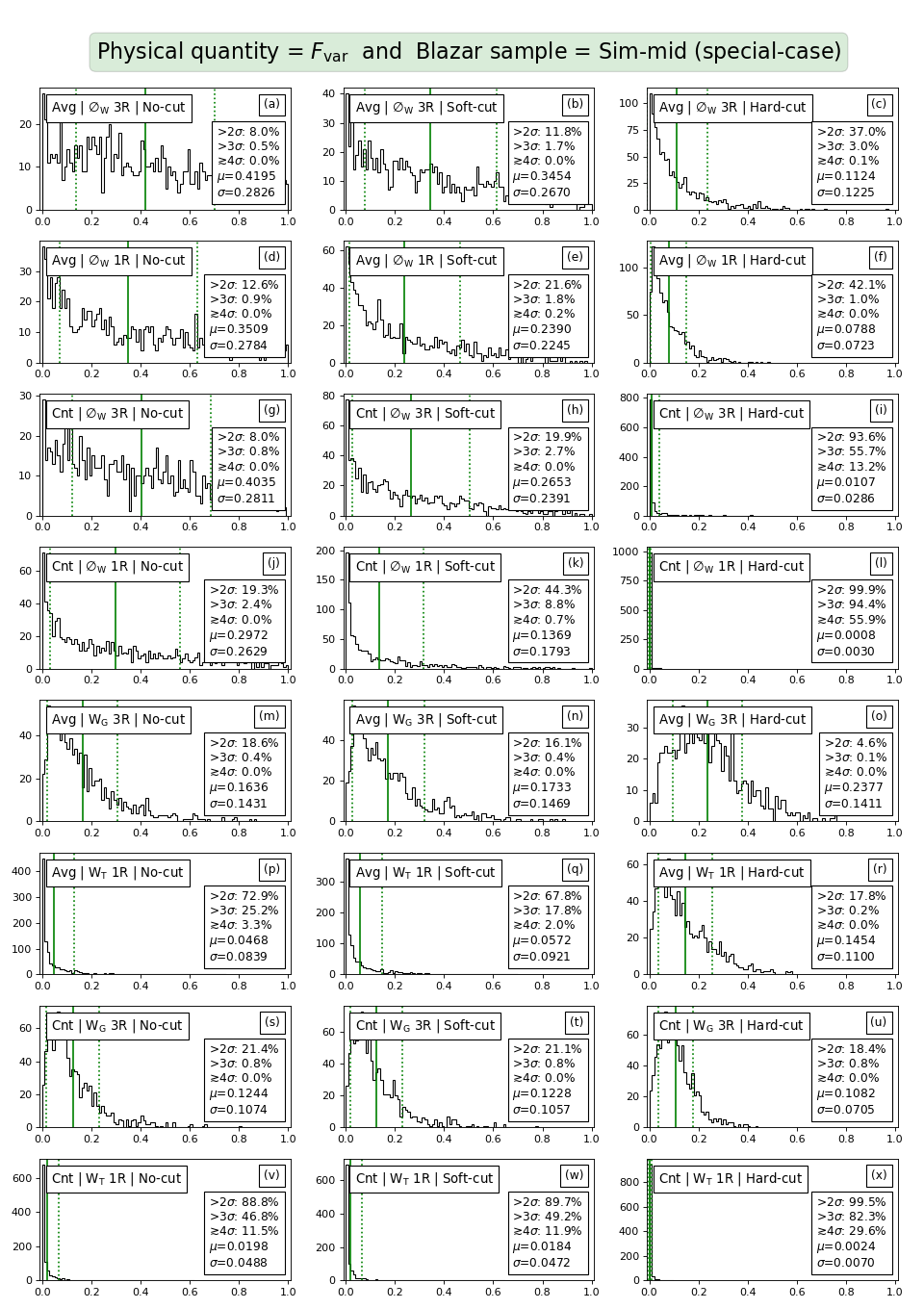}
    \caption{P-value distributions when using the time-averaged $F_\mathrm{var}$ and the sim-mid blazars. For more details, see the text in Appendix \ref{appendix_pval_dist}.}
    \label{fig_pval_histograms_sim_Mid_Fvar}
\end{figure*}

\begin{figure*}
    \centering
    \includegraphics[width=17cm, keepaspectratio]{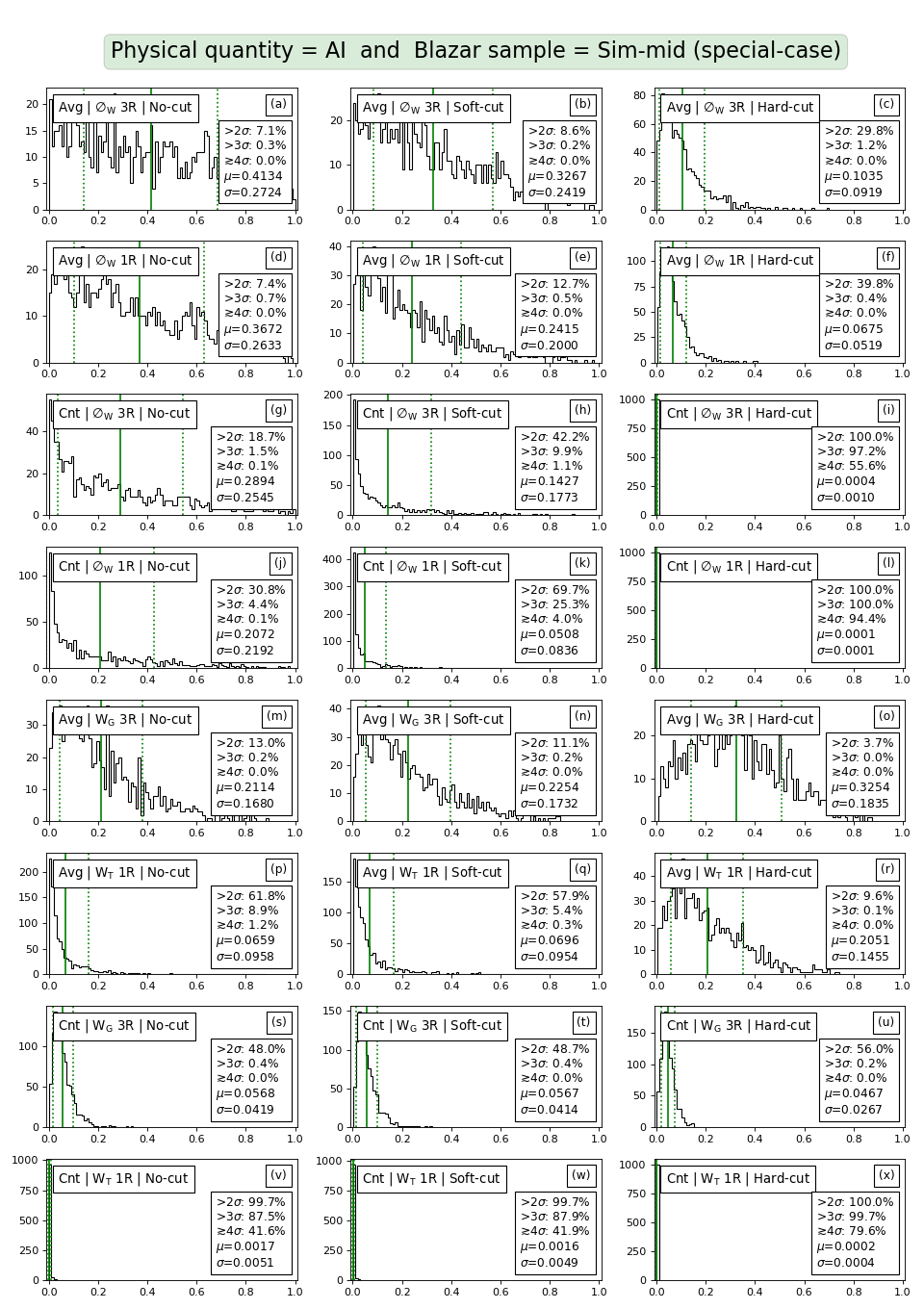}
    \caption{P-value distributions when using the time-resolved AI and the sim-mid blazars. For more details, see the text in Appendix \ref{appendix_pval_dist}.}
    \label{fig_pval_histograms_sim_Mid_AI}
\end{figure*}

\begin{figure*}
    \centering
    \includegraphics[width=17cm, keepaspectratio]{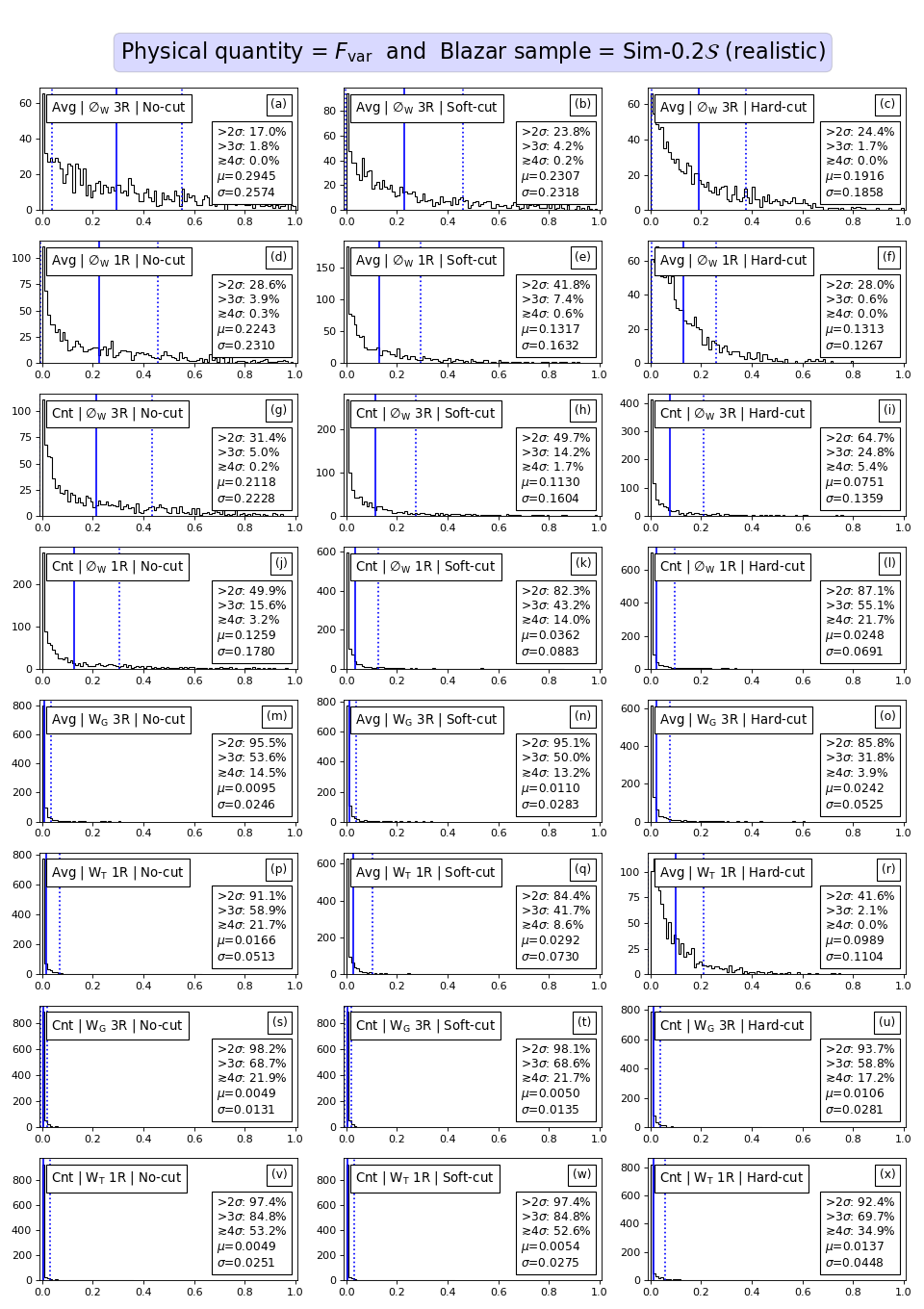}
    \caption{P-value distributions when using the time-averaged $F_\mathrm{var}$ and the sim-$0.2\mathcal{S}$ blazars. For more details, see the text in Appendix \ref{appendix_pval_dist}.}
    \label{fig_pval_histograms_sim_Sness_Mask0.2_Fvar}
\end{figure*}

\begin{figure*}
    \centering
    \includegraphics[width=17cm, keepaspectratio]{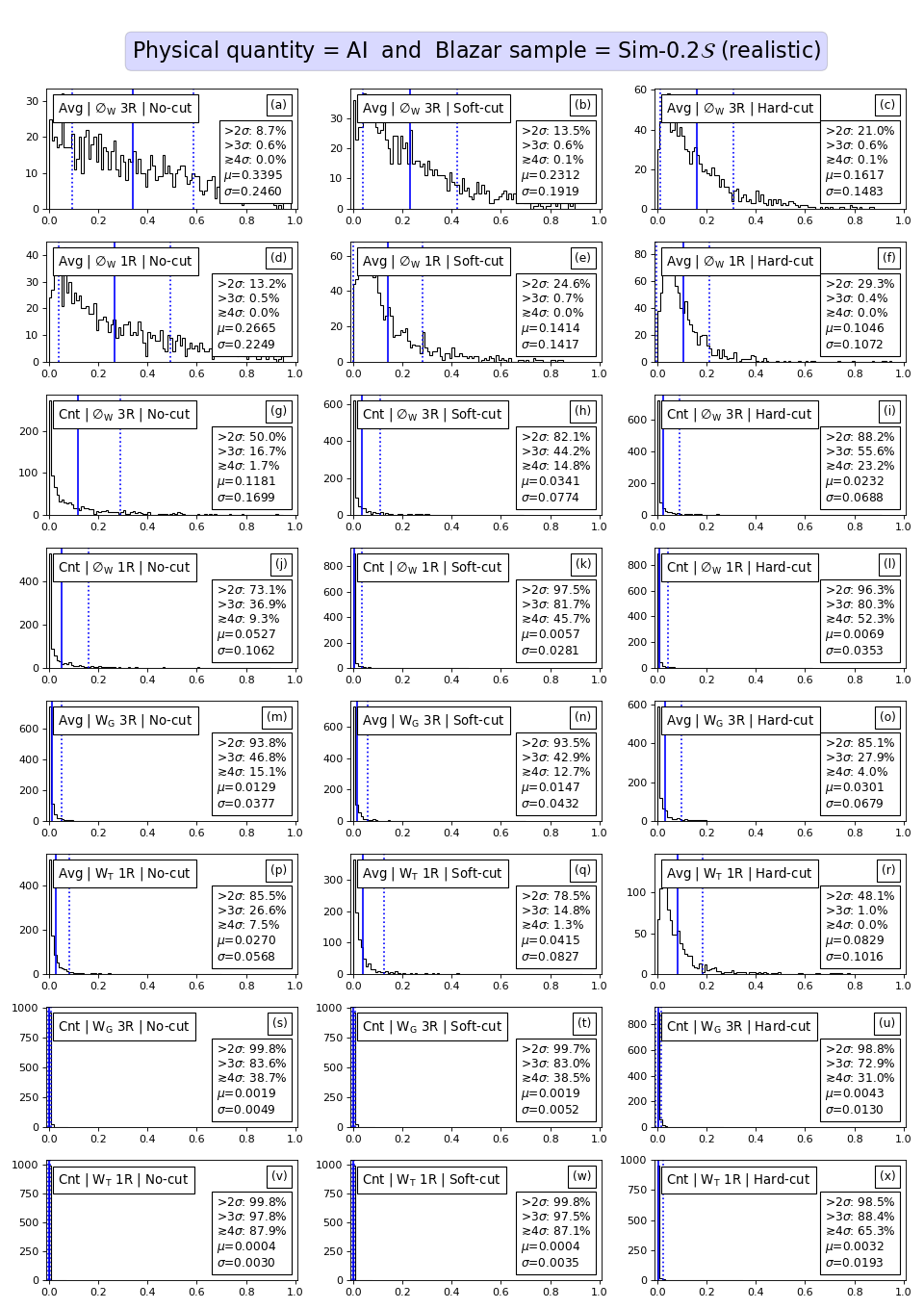}
    \caption{P-value distributions when using the time-resolved AI and the sim-$0.2\mathcal{S}$ blazars. For more details, see the text in Appendix \ref{appendix_pval_dist}.}
    \label{fig_pval_histograms_sim_Sness_Mask0.2_AI}
\end{figure*}

\begin{figure*}
    \centering
    \includegraphics[width=17cm, keepaspectratio]{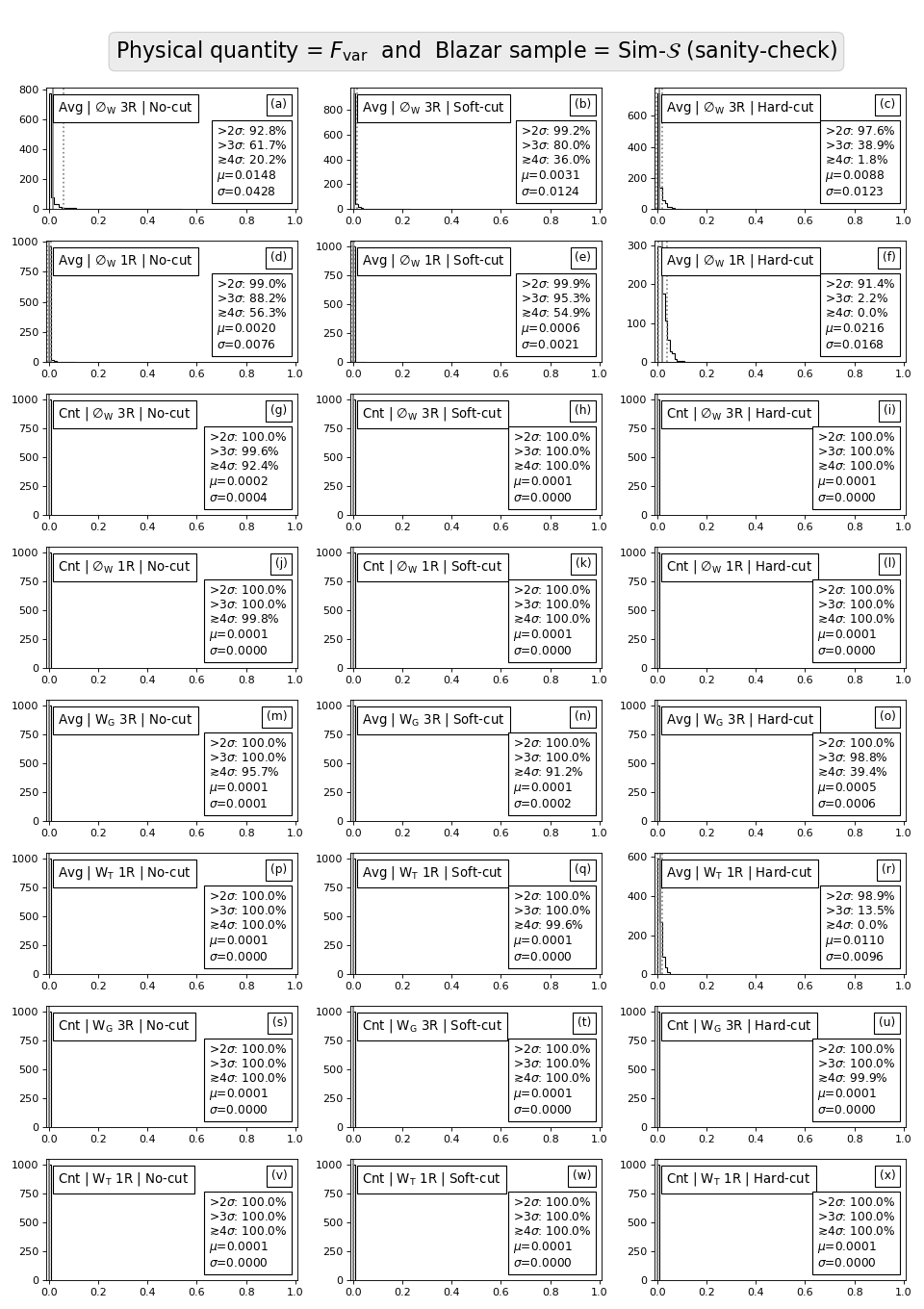}
    \caption{P-value distributions when using the time-averaged $F_\mathrm{var}$ and the sim-$\mathcal{S}$ blazars. For more details, see the text in Appendix \ref{appendix_pval_dist}.}
    \label{fig_pval_histograms_sim_Sness_Fvar}
\end{figure*}

\begin{figure*}
    \centering
    \includegraphics[width=17cm, keepaspectratio]{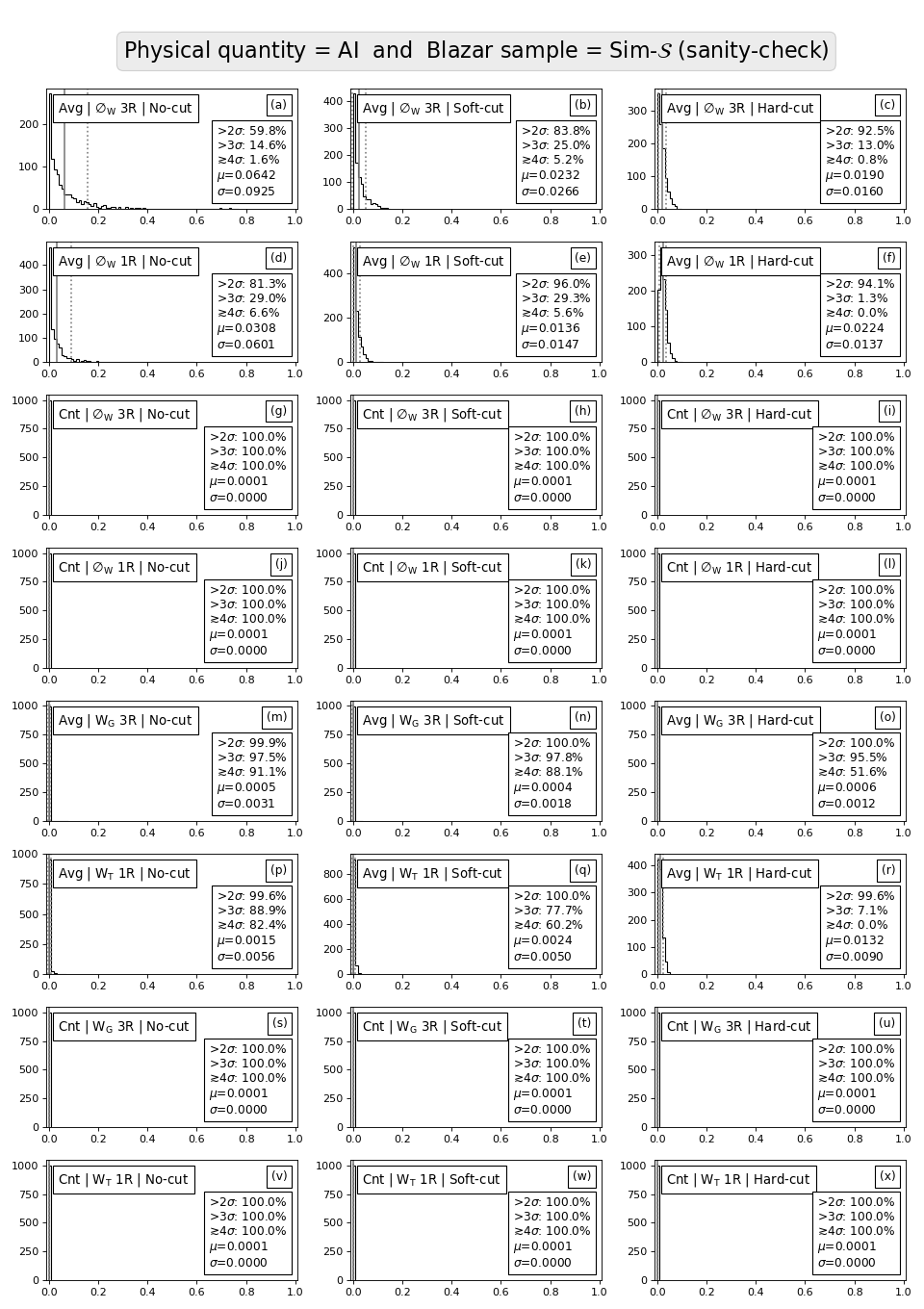}
    \caption{P-value distributions when using the time-resolved AI and the sim-$\mathcal{S}$ blazars. For more details, see the text in Appendix \ref{appendix_pval_dist}.}
    \label{fig_pval_histograms_sim_Sness_AI}
\end{figure*}

Figures \ref{fig_pval_histograms_sim_Null_Fvar} to \ref{fig_pval_histograms_sim_Sness_AI} show all the 240 p-value distributions of this study. These are divided into ten figures. At the top of each figure, a colored text card shows the variability measure and the blazar sample used in obtaining their respective p-value distributions. The odd numbered figures (\ref{fig_pval_histograms_sim_Null_Fvar}, \ref{fig_pval_histograms_sim_Best_Fvar}, \ref{fig_pval_histograms_sim_Mid_Fvar}, \ref{fig_pval_histograms_sim_Sness_Mask0.2_Fvar}, and \ref{fig_pval_histograms_sim_Sness_Fvar}) use $F_\mathrm{var}$ as the variability measure, while even numbered figures (\ref{fig_pval_histograms_sim_Null_AI}, \ref{fig_pval_histograms_sim_Best_AI}, \ref{fig_pval_histograms_sim_Mid_AI}, \ref{fig_pval_histograms_sim_Sness_Mask0.2_AI}, and \ref{fig_pval_histograms_sim_Sness_AI}) use AI. The blazar samples are as follows:
\begin{itemize}
    \item Sim-null sample (purple text card; Figs. \ref{fig_pval_histograms_sim_Null_Fvar} and \ref{fig_pval_histograms_sim_Null_AI})
    \item Sim-best sample (orange text card; Figs. \ref{fig_pval_histograms_sim_Best_Fvar} and \ref{fig_pval_histograms_sim_Best_AI})
    \item Sim-mid sample (green text card; Figs. \ref{fig_pval_histograms_sim_Mid_Fvar} and \ref{fig_pval_histograms_sim_Mid_AI})
    \item Sim-$0.2\mathcal{S}$ sample (blue text card; Figs. \ref{fig_pval_histograms_sim_Sness_Mask0.2_Fvar} and \ref{fig_pval_histograms_sim_Sness_Mask0.2_AI})
    \item Sim-$\mathcal{S}$ sample (gray text card; Figs. \ref{fig_pval_histograms_sim_Sness_Fvar} and \ref{fig_pval_histograms_sim_Sness_AI})
\end{itemize}

In each of the ten figures, there are 24 subplots. These are labeled from (a) to (x) as shown in their top right text box. Each of these subplots shows the p-value distribution of a specific test strategy, which is labeled by the text in the top left box (more details below). The bottom right text box in each subplot gives information about the statistics of the p-value distribution: >2$\sigma$, >3$\sigma$, and $\gtrsim$4$\sigma$ refer to the fraction (in \%) of p-values that cross the $2\sigma$ ($p<0.0455$), $3\sigma$ ($p<0.0027$), and $\sim$4$\sigma$ ($p=0.0001$; the smallest possible p-value obtainable from our tests) thresholds, respectively. We note that the fraction labeled as >3$\sigma$ is the same as $\mathcal{F}_{3\sigma}$, which is also reported in Table \ref{results_table} and used in the discussions of Sect. \ref{sec_results}. The last two entries in the bottom right text box of each subplot are $\mu$ and $\sigma$, which are the mean and standard deviation of the p-value distributions, respectively. $\mu$ and $\sigma$ are visualized on each distribution using vertical lines (the solid line is $\mu$ and the dotted lines are $\mu \pm \sigma$, as long as mathematically appropriate, i.e., $0 \leq \mu \pm \sigma \leq 1$). 

As mentioned above, the top left text box contains information about the specific test strategy to which the distribution pertains. The neutrino sample cut (see Sect. \ref{sec_tests_neut_cuts}) results are shown as follows:
\begin{itemize}
    \item No-cut sample: left subplots (a, d, g, j, m, p, s, and v)
    \item Soft-cut sample: middle subplots (b, e, h, k, n, q, t, and w)
    \item Hard-cut sample: right subplots (c, f, i, l, o, r, u, and x)
\end{itemize}
The unweighted and weighted test (see Sect. \ref{sec_tests_weighting}) results are shown as follows:
\begin{itemize}
    \item $\varnothing_\mathrm{W}$ 3R: rows one and three; subplots (a, b, c, g, h, and i)
    \item $\varnothing_\mathrm{W}$ 1R: rows two and four; subplots (d, e, f, j, k, and l)
    \item $\mathrm{W}_\mathrm{G}$ 3R: rows five and seven; subplots (m, n, o, s, t, and u)
    \item $\mathrm{W}_\mathrm{T}$ 1R: rows six and eight; subplots (p, q, r, v, w, and x)
\end{itemize}
Results of the averaging- and counting-based tests (denoted as "Avg" and "Cnt", respectively; see Sect. \ref{sec_tests_TS_parameters}) are as follows:
\begin{itemize}
    \item Avg: rows one, two, five, and six; subplots (a--f and m--r)
    \item Cnt: rows three, four, seven, and eight; subplots (g--l and s--x)
\end{itemize}

Regarding the general shape and behavior of the p-value distributions, it is apparent that most of them are well behaved apart from a few oddities that we discuss below. Firstly, from the p-value distributions in the sim-null case (Figs. \ref{fig_pval_histograms_sim_Null_Fvar} and \ref{fig_pval_histograms_sim_Null_AI}), we see that almost all the p-values are nearly uniformly spread across $0 \le p \le 1$ with the mean p-value ($\mu$) being around 0.5 and the standard deviation ($\sigma$) being around 0.3. Moreover, the fractions represented by >2$\sigma$ and >3$\sigma$ are around 5\% and 0.3\%, respectively. These clearly demonstrate that most of the p-value distributions in the sim-null case are near-Gaussian random distributions --- as one would expect from such a random process. However, there are a few nonuniform distributions, namely those seen in subplots (l) and (x) of Fig. \ref{fig_pval_histograms_sim_Null_AI}. These occur when counting those AI values that surpass a predetermined threshold. 

The most noticeable nonuniform feature is the peak of p-values at $p=1$. This is an artifact of how Eq. \ref{eqn_pval} is defined: when the observed TS value is exactly zero, $M=N$, which leads to $p=1$. It turns out that counting AI values that surpass a threshold as well as setting a hard limit on the sample of neutrinos leads to many simulations where the global TS parameter ends up being a small, whole number (e.g., 0, 1, 2, etc.). This is because the hard-cut subsample of neutrinos ignores 75\% of the whole IceCat1+ neutrino sample and only consists of 71 spatially well-constrained neutrinos, which drastically reduces the number of blazar--neutrino spatial associations in the 1R test strategies (for both $\varnothing_\mathrm{W}$ and $\mathrm{W}_\mathrm{T}$). Additionally, when using the counted TS parameters, only those associations with AI above a threshold contribute to the global TS parameter. Unfortunately, Eq. \ref{eqn_pval} does not perform ideally under such a circumstance when the TS parameters are small, whole numbers, leading to clumping of p-values at specific intervals. However, while these p-value distributions do exhibit nonuniformity and an overall abnormal shape (as compared to the rest of the sim-null distributions), they do not result in smaller than average p-values. In fact, the nonuniformity appears to push these p-value distributions to the higher end (i.e., $p \rightarrow 1$), possibly increasing the number of false-negatives; critically, we emphasize that the number of false-positives remains unaffected. Nevertheless, this is a caveat of setting a hard cut on the neutrino sample, especially when combined with a counting-based strategy that leads to small, whole number TS parameters. Interestingly, in the weighted scenarios, the distribution is more uniform, which can be interpreted as yet another reason why weighting is better than setting neutrino sample cuts. Finally, we point out that such p-value clumping is also somewhat present in the sim-best case (with no weighting employed) when setting a hard-cut on the neutrinos and counting the AI (see subplot l of Fig. \ref{fig_pval_histograms_sim_Best_AI}) for a similar reason as above, since its TS parameters are also small, whole numbers.

\end{appendix}

\end{document}